\documentclass[proof]{WileyASNA-v1}

\articletype{Original Paper}%

\received{<day> <Month>, <year>}
\revised{<day> <Month>, <year>}
\accepted{<day> <Month>, <year>}

\begin{document}

\title{Red quasars: estimation of SMBH spin, mass and accretion disk inclination angle \\ {\normalsize \it Accepted for publication in Astronomische Nachrichten}}

\author[1]{M. Yu. Piotrovich*}

\author[1]{S. D. Buliga}

\author[1]{T. M. Natsvlishvili}

\authormark{PIOTROVICH \textsc{et al}}

\address[1]{\orgname{Central Astronomical Observatory at Pulkovo}, \orgaddress{\state{St.-Petersburg}, \country{Russia}}}

\corres{*M.Yu. Piotrovich, Central Astronomical Observatory at Pulkovo, St.-Petersburg, Russia. \email{mpiotrovich@mail.ru}}

\abstract{We estimated values of spin, mass and inclination angle for sample of 42 red quasars. Our estimations show that 2 objects: F2MS~J1113+1244 and F2MS~J1434+0935 with highest Eddington ratios may have geometrically thick disk. Six objects: SDSS~J0036-0113, S82X~0040+0058, S82X~0118+0018, S82X~0303-0115, FBQS~J1227+3214, S82X~2328-0028 may have ''retrograde'' rotation. Analysis of estimated spin values shows that red quasar population may contain Seyfert galaxies and NLS1.}


\jnlcitation{\cname{%
\author{M.Yu. Piotrovich},
\author{S.D. Buliga}, and
\author{T.M. Natsvlishvili}} (\cyear{2024}),
\ctitle{Red quasars: estimation of SMBH spin, mass and accretion disk inclination angle},
\cjournal{Astronomische Nachrichten}, \cvol{2024;00:1--6}.}

\maketitle

\section{Introduction}

Large ground-based optical surveys, such as the Sloan Digital Sky Survey \citep{lyke20}, have detected more than three-quarters of a million active galactic nuclei (AGNs). Some of these objects are sources in which we can directly observe the central object, which makes it easy to identify. However, there are darkened star systems in which the optical view of the central engine is obscured by a circumnuclear torus (Type 2 AGN) \citep{antonucci93,urry95} and large amounts of dust from the host galaxy. Red quasars are typically AGNs with broad emission lines, but are also covered in large amounts of dust, which reddens the spectrum and attenuates the optical emission. Studies of this extreme segment of the darkened AGN population indicate that their red color is due to the stage of AGN evolution in the black hole growth model of a major galaxy mergers, rather than to the orientation of the dust torus relative to the line of sight \citep{georgakakis09,glikman12}. According to this model, the obscuring dust is eventually dispersed by the quasar's powerful winds, revealing the quasar's luminous core. In this scenario, dust-reddened (or ''red'') quasars represent an early phase of supermassive black hole (SMBH)/galaxy coevolution: the transition from a dust-covered core to a typical unobscured quasar \citep{lamassa17}.

Determining the spin (dimensionless angular momentum) value $a = cJ/(G M_\text{BH}^2)$ (where $J$ is the angular momentum, $M_\text{BH}$ is the black hole mass, and $c$ is the speed of light) of the SMBH at the centre of an AGN is a crucial problem in modern astrophysics. The spin is thought to play a pivotal role in generating relativistic jets in AGNs; consequently, it is often the power of these relativistic jets that is used to infer the spin of the SMBH \citep{daly11}. Typically, the kinetic power of a relativistic jet is estimated by determining the magnetic field strength near the SMBH event horizon through various mechanisms, including the Blandford-Znajek mechanism \citep{blandford77}, the Blandford-Payne mechanism \citep{blandford82}, and the Garofalo mechanism \citep{garofalo10}. However in this work, we will determine spin using a different method - through radiative efficiency of accretion disk, which strongly depends on spin \citep{bardeen72,novikov73,krolik07,krolik07b}.

\setlength{\tabcolsep}{5pt}
\begin{table*}
\caption{Physical parameters of our objects from literature: cosmological redshift $z$, bolometric luminosity $L_{\rm bol}$ (in $\log(L_{\rm bol}[{\rm erg/s}])$), Eddington ratio $l_{\rm E}$ and SMBH mass $M_{\rm BH}$ (in $\log(M_{\rm BH}/M_\odot)$), and estimated values of inclination angle $i_{1,2,3}$ (in [deg]), mass $M_{\rm BH,1,2,3}$ and spin $a_{1,2,3}$ for 3 models: (1) \citet{raimundo11}, (2) \citet{trakhtenbrot14}, (3) \citet{du14}.}
\centering
\begin{tabular}{l|cccc|ccc|ccc|ccc}
\hline
Object & $\log{z}$ & $L_\text{bol}$ & $\log{l_\text{E}}$ & $M_\text{BH}$ & $i_1$ & $M_\text{BH,1}$ & $a_1$ & $i_2$ & $M_\text{BH,2}$ & $a_2$ & $i_3$ & $M_\text{BH,3}$ & $a_3$\\
\hline
S82X  0011+0057 &  0.173 & 46.02$^{3}$ & -0.96$^{3}$ & 8.85$^{3}$ & 60 & 8.73 &  0.982 & 45 & 8.91 &  0.968 & 50 & 8.84 &  0.998\\
S82X  0022+0020 & -0.097 & 45.85$^{3}$ & -1.00$^{2}$ & 8.75$^{3}$ & 50 & 8.74 &  0.998 & 45 & 8.81 &  0.970 & 50 & 8.74 &  0.998\\
SDSS J0036-0113 & -0.081 & 45.79$^{7}$ & -0.08$^{7}$ & 7.78$^{7}$ & 35 & 7.56 & -0.506 & 25 & 7.83 & -0.452 & 30 & 7.68 & -0.506\\
S82X  0040+0058 & -0.091 & 45.12$^{3}$ & -0.29$^{0}$ & 7.30$^{3}$ & 45 & 7.36 & -0.354 & 40 & 7.44 & -0.754 & 45 & 7.36 & -0.824\\
S82X  0043+0052 & -0.082 & 45.93$^{3}$ & -1.05$^{3}$ & 8.84$^{3}$ & 60 & 8.72 &  0.990 & 45 & 8.90 &  0.980 & 55 & 8.77 &  0.986\\
S82X  0100+0008 &  0.173 & 45.52$^{3}$ & -1.00$^{3}$ & 8.43$^{3}$ & 45 & 8.01 &  0.644 & 45 & 8.01 &  0.232 & 45 & 8.01 &  0.682\\
S82X  0118+0018 &  0.043 & 45.41$^{3}$ & -0.69$^{0}$ & 7.99$^{3}$ & 45 & 7.57 & -0.824 & 40 & 7.65 & -0.686 & 45 & 7.57 & -0.624\\
S82X  0242+0005 &  0.394 & 47.24$^{3}$ & -0.49$^{3}$ & 9.62$^{3}$ & 50 & 9.45 &  0.998 & 45 & 9.52 &  0.890 & 50 & 9.45 &  0.990\\
S82X  0302-0003 &  0.099 & 46.85$^{3}$ & -0.64$^{3}$ & 9.38$^{3}$ & 45 & 9.20 &  0.998 & 45 & 9.20 &  0.828 & 45 & 9.20 &  0.994\\
S82X  0303-0115 & -0.228 & 45.17$^{3}$ & -0.39$^{3}$ & 7.44$^{3}$ & 25 & 7.44 & -0.038 & 25 & 7.44 & -0.354 & 30 & 7.29 & -0.624\\
 F2M J0729+3336 & -0.019 & 46.91$^{4}$ &  0.10$^{4}$ & 8.70$^{4}$ & 45 & 8.74 &  0.882 & 45 & 8.74 & -0.354 & 45 & 8.74 &  0.598\\
F2MS J0825+4716 & -0.095 & 47.35$^{2}$ & -0.17$^{2}$ & 9.10$^{5}$ & 45 & 8.99 &  0.900 & 45 & 8.99 & -0.506 & 45 & 8.99 &  0.608\\
 F2M J0830+3759 & -0.383 & 45.59$^{6}$ & -0.40$^{4}$ & 8.60$^{4}$ & 45 & 8.05 &  0.852 & 45 & 8.05 &  0.186 & 45 & 8.05 &  0.676\\
F2MS J0834+3506 & -0.328 & 46.18$^{2}$ & -1.40$^{2}$ & 9.50$^{2}$ & 70 & 9.11 &  0.992 & 50 & 9.29 &  0.998 & 65 & 9.14 &  0.990\\
F2MS J0841+3604 & -0.257 & 46.21$^{4}$ & -0.40$^{4}$ & 8.50$^{4}$ & 45 & 8.40 &  0.884 & 45 & 8.40 & -0.004 & 45 & 8.40 &  0.668\\
F2MS J0911+0143 & -0.220 & 46.19$^{7}$ & -0.31$^{7}$ & 8.40$^{7}$ & 45 & 8.46 &  0.934 & 45 & 8.46 &  0.254 & 45 & 8.46 &  0.778\\
F2MS J0915+2418 & -0.074 & 47.70$^{2}$ & -0.03$^{2}$ & 9.00$^{7}$ & 45 & 9.06 &  0.804 & 40 & 9.14 & -0.506 & 45 & 9.06 &  0.332\\
F2MS J1012+2825 & -0.028 & 47.31$^{4}$ & -0.10$^{4}$ & 9.30$^{4}$ & 50 & 9.29 &  0.994 & 45 & 9.36 &  0.682 & 45 & 9.36 &  0.978\\
 F2M J1106+4807 & -0.362 & 46.73$^{2}$ & -0.28$^{2}$ & 8.90$^{2}$ & 45 & 8.96 &  0.996 & 45 & 8.96 &  0.618 & 45 & 8.96 &  0.946\\
F2MS J1113+1244 & -0.167 & 47.48$^{2}$ &  0.36$^{2}$ & 8.50$^{7}$ & 45 & 9.36 &  0.998 & 45 & 9.36 &  0.500 & 45 & 9.36 &  0.946\\
F2MS J1118-0033 & -0.164 & 46.94$^{2}$ &  0.04$^{2}$ & 8.80$^{2}$ & 45 & 8.86 &  0.946 & 45 & 8.86 &  0.026 & 45 & 8.86 &  0.762\\
F2MS J1151+5359 & -0.108 & 46.57$^{2}$ & -0.30$^{2}$ & 8.80$^{2}$ & 45 & 8.86 &  0.994 & 45 & 8.86 &  0.618 & 45 & 8.86 &  0.940\\
SDSS J1209-0107 & -0.442 & 46.51$^{7}$ & -0.52$^{7}$ & 8.90$^{7}$ & 50 & 8.89 &  0.994 & 45 & 8.96 &  0.804 & 45 & 8.96 &  0.986\\
FBQS J1227+3214 & -0.863 & 45.55$^{2}$ & -0.10$^{2}$ & 7.40$^{1}$ & 40 & 7.54 & -0.220 & 30 & 7.76 & -0.220 & 35 & 7.64 & -0.180\\
F2MS J1227+5053 & -0.116 & 46.35$^{7}$ & -0.19$^{7}$ & 8.50$^{7}$ & 45 & 8.67 &  0.982 & 45 & 8.67 &  0.524 & 45 & 8.67 &  0.902\\
F2MS J1248+0531 & -0.126 & 46.28$^{7}$ & -0.20$^{7}$ & 8.40$^{7}$ & 45 & 8.46 &  0.906 & 45 & 8.46 &  0.084 & 45 & 8.46 &  0.710\\
F2MS J1307+2338 & -0.561 & 45.51$^{7}$ & -0.28$^{7}$ & 7.70$^{7}$ & 45 & 7.76 &  0.430 & 45 & 7.76 & -0.824 & 45 & 7.76 &  0.026\\
SDSS J1309+6042 & -0.193 & 46.26$^{7}$ & -0.21$^{7}$ & 8.40$^{7}$ & 45 & 8.09 &  0.294 & 35 & 8.27 & -0.180 & 45 & 8.09 & -0.452\\
F2MS J1313+1453 & -0.234 & 46.67$^{7}$ & -0.30$^{7}$ & 8.90$^{7}$ & 45 & 8.96 &  0.998 & 45 & 8.96 &  0.676 & 45 & 8.96 &  0.960\\
 F2M J1324+0537 & -0.688 & 46.72$^{2}$ & -0.12$^{2}$ & 8.70$^{2}$ & 45 & 8.76 &  0.950 & 45 & 8.76 &  0.162 & 45 & 8.76 &  0.788\\
F2MS J1434+0935 & -0.114 & 46.59$^{7}$ &  0.53$^{7}$ & 8.00$^{7}$ & 45 & 8.70 &  0.952 & 45 & 8.70 &  0.232 & 45 & 8.70 &  0.804\\
 F2M J1507+3129 & -0.005 & 46.93$^{2}$ &  0.17$^{2}$ & 8.70$^{2}$ & 45 & 8.76 &  0.886 & 45 & 8.76 & -0.354 & 45 & 8.76 &  0.608\\
 F2M J1531+2423 &  0.359 & 48.21$^{2}$ &  0.21$^{2}$ & 9.90$^{2}$ & 55 & 9.83 &  0.996 & 45 & 9.96 &  0.758 & 45 & 9.96 &  0.998\\
F2MS J1532+2415 & -0.249 & 46.57$^{2}$ & -0.54$^{2}$ & 8.80$^{2}$ & 45 & 8.86 &  0.994 & 45 & 8.86 &  0.618 & 45 & 8.86 &  0.940\\
F2MS J1540+4923 & -0.157 & 46.44$^{7}$ & -0.72$^{7}$ & 9.10$^{7}$ & 60 & 8.98 &  0.990 & 45 & 9.16 &  0.962 & 50 & 9.09 &  0.998\\
F2MS J1600+3522 & -0.151 & 46.63$^{7}$ &  0.29$^{7}$ & 8.30$^{7}$ & 45 & 8.36 &  0.524 & 40 & 8.44 & -0.686 & 45 & 8.36 & -0.106\\
F2MS J1656+3821 & -0.135 & 46.81$^{2}$ & -0.12$^{2}$ & 8.70$^{7}$ & 45 & 8.76 &  0.928 & 45 & 8.76 & -0.038 & 45 & 8.76 &  0.724\\
 F2M J1715+2807 & -0.281 & 46.55$^{2}$ & -0.48$^{2}$ & 8.90$^{2}$ & 50 & 8.89 &  0.992 & 45 & 8.96 &  0.778 & 45 & 8.96 &  0.982\\
F2MS J1720+6156 & -0.138 & 46.41$^{7}$ &  0.21$^{7}$ & 8.10$^{7}$ & 45 & 8.39 &  0.778 & 45 & 8.39 & -0.564 & 45 & 8.39 &  0.416\\
F2MS J2325-1052 & -0.249 & 46.31$^{7}$ & -0.21$^{7}$ & 8.40$^{7}$ & 45 & 8.72 &  0.994 & 45 & 8.72 &  0.652 & 45 & 8.72 &  0.942\\
S82X  2328-0028 & -0.232 & 45.08$^{3}$ & -0.38$^{3}$ & 7.34$^{3}$ & 45 & 7.40 & -0.824 & 40 & 7.48 & -0.402 & 45 & 7.40 & -0.452\\
F2MS J2339-0912 & -0.180 & 46.90$^{7}$ &  0.04$^{7}$ & 8.80$^{7}$ & 45 & 8.50 &  0.548 & 40 & 8.58 & -0.824 & 45 & 8.50 & -0.142\\
\hline
\multicolumn{14}{l}{(0) Our calculations; (1) \citet{glikman17};}\\
\multicolumn{14}{l}{(2) \citet{glikman24}; (3) \citet{lamassa17}; (4) \citet{urrutia12};}\\
\multicolumn{14}{l}{(5) \citet{kim15}; (6) \citet{lamassa16}; (7) \citet{kim18}.}\\
\hline
\end{tabular}
\label{tab1}
\end{table*}

\begin{figure*}[!htbp]
\centering
\includegraphics[bb= 50 0 715 525, clip, width=1.0 \columnwidth]{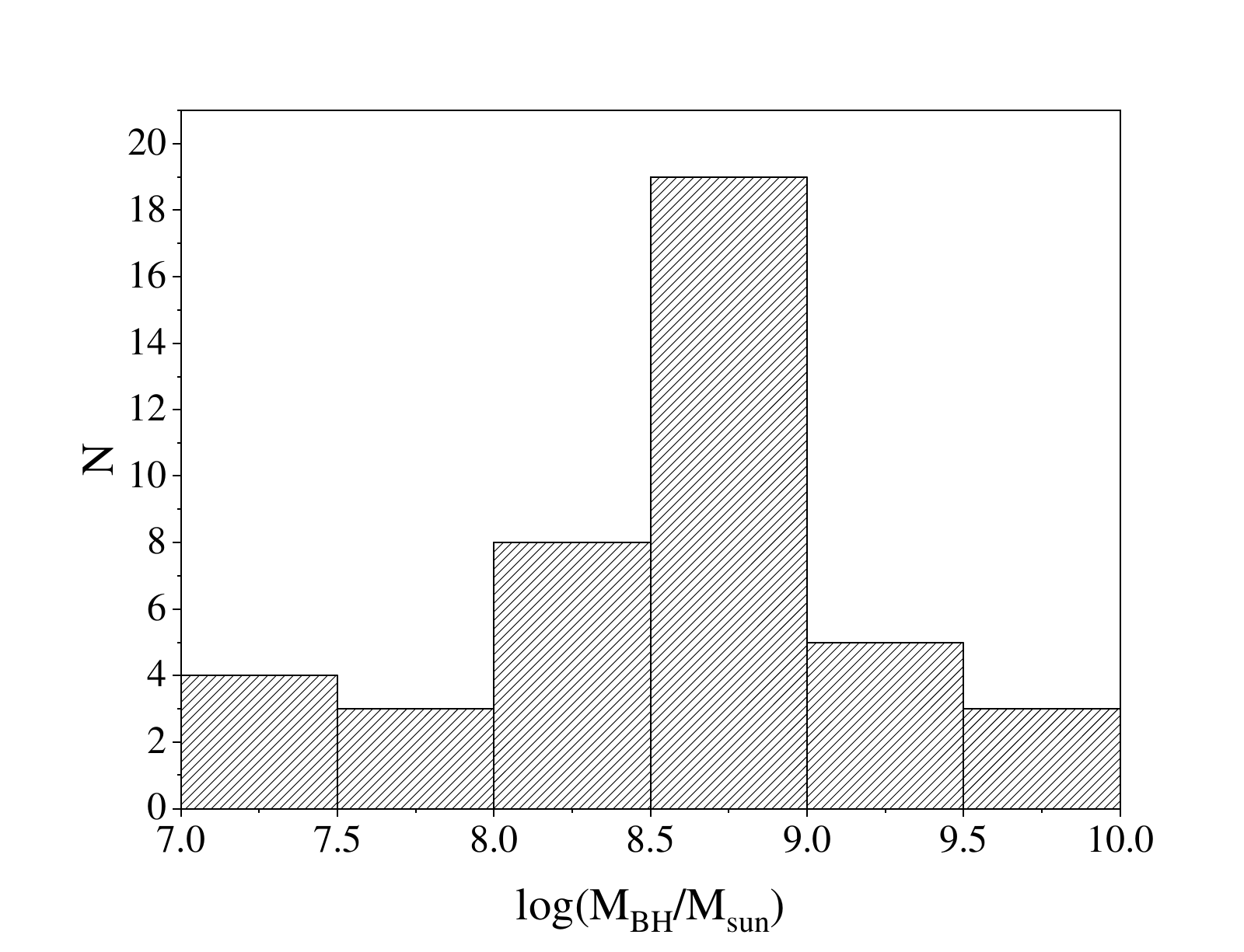}
\includegraphics[bb= 50 0 715 530, clip, width=1.0 \columnwidth]{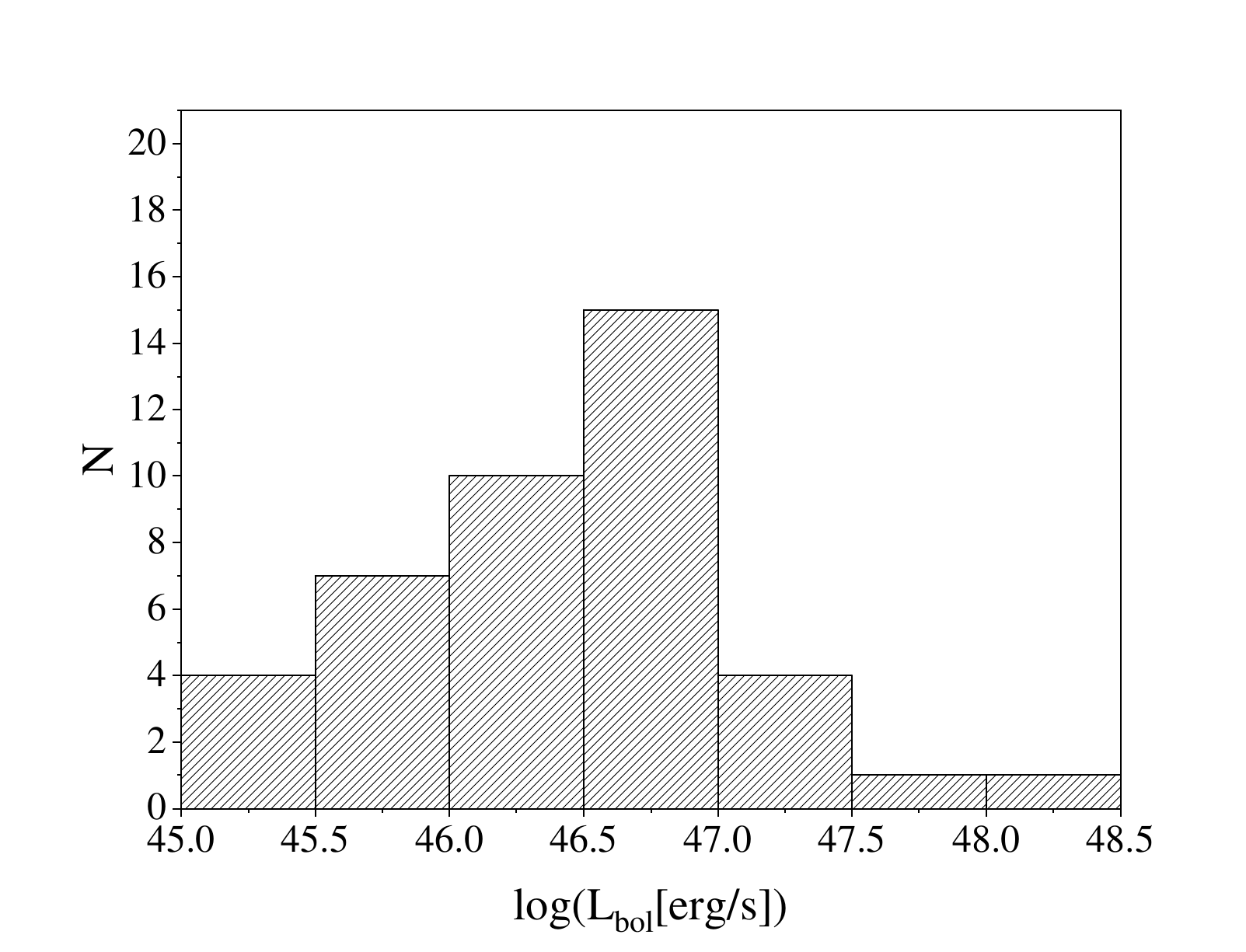}
\includegraphics[bb= 50 0 715 530, clip, width=1.0 \columnwidth]{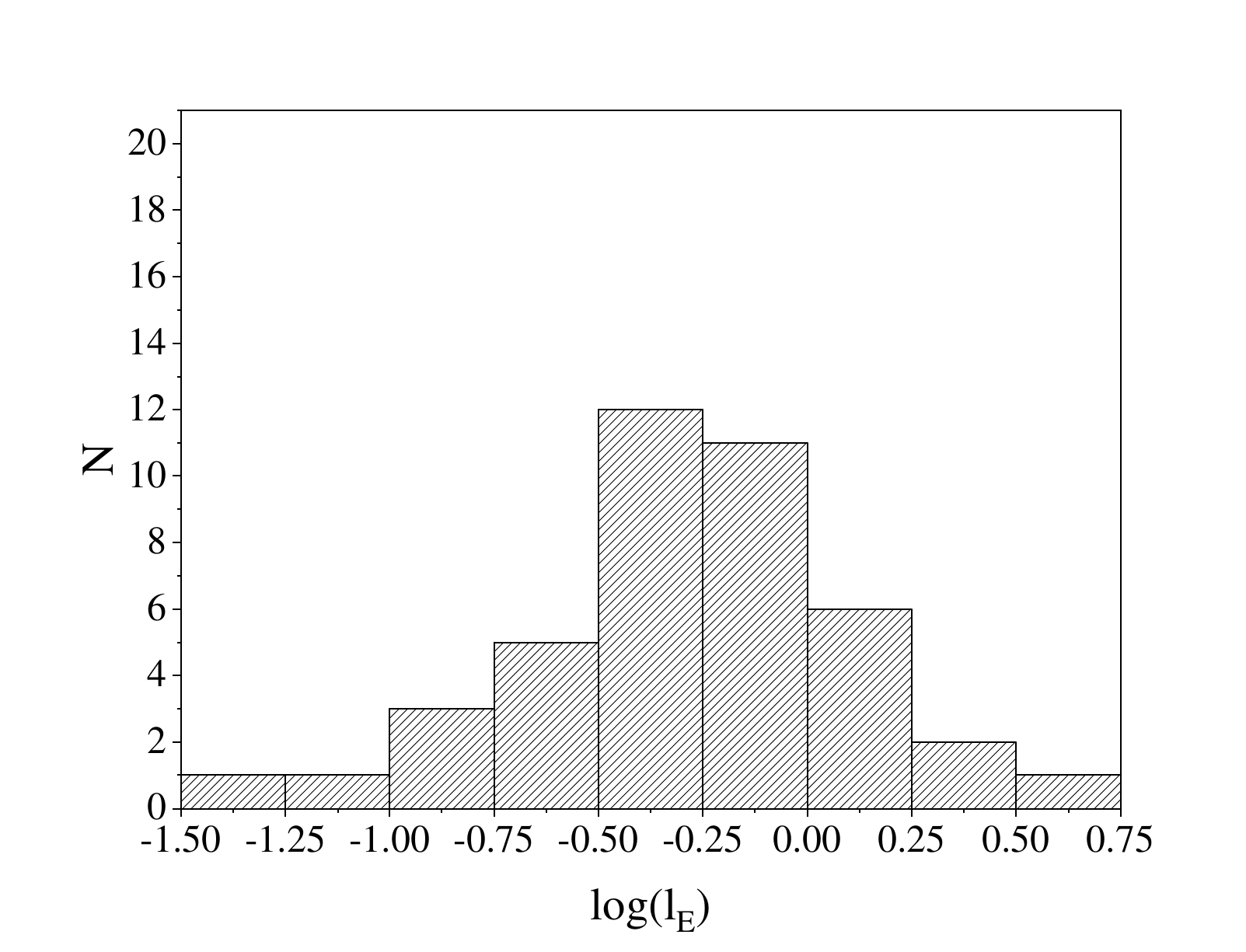}
\includegraphics[bb= 50 0 715 525, clip, width=1.0 \columnwidth]{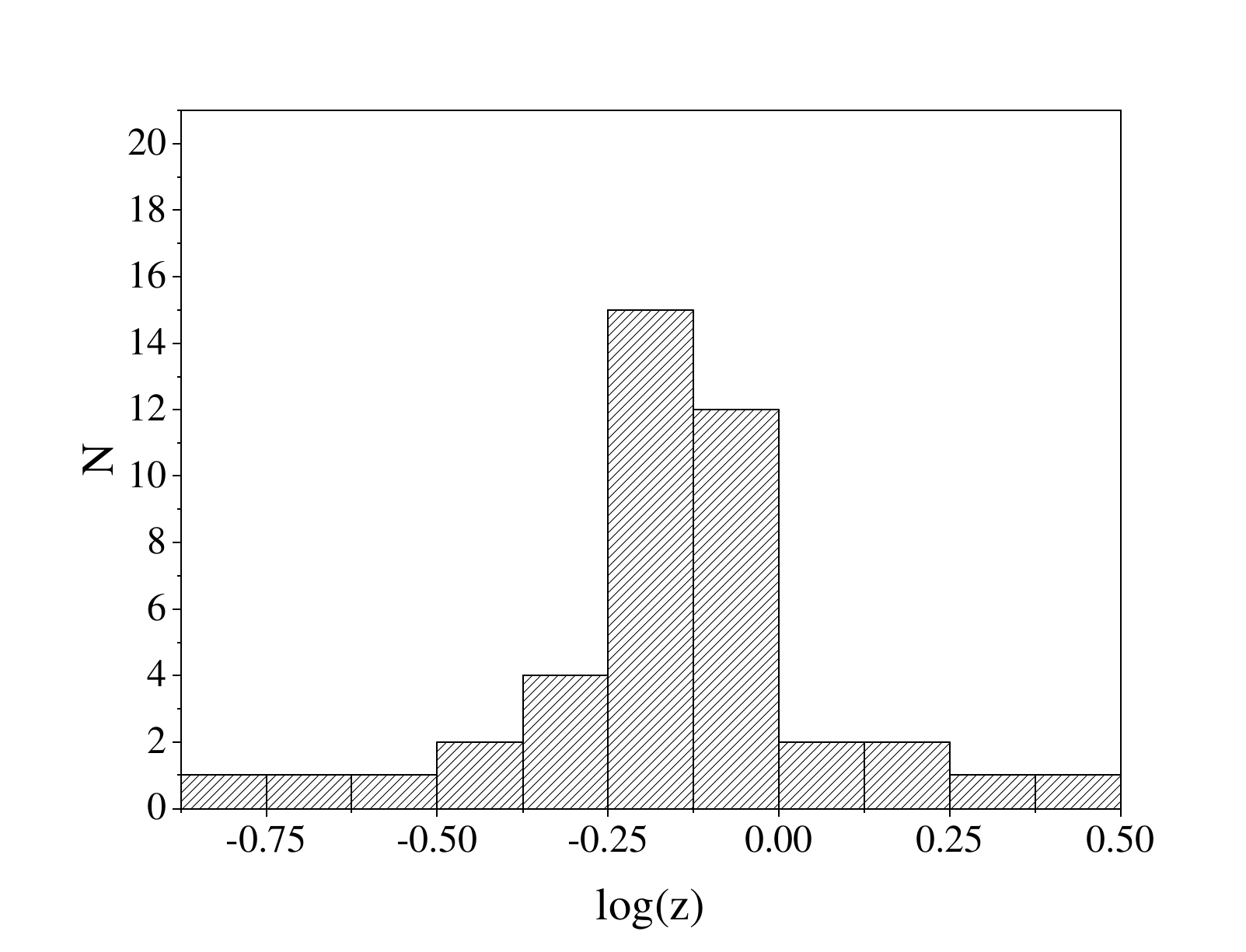}
\caption{Distributions of the parameters of objects from the initial set: SMBH mass $M_\text{BH}$, bolometric luminosity $L_\text{bol}$, Eddington ratio $l_\text{E}$ and redshift $z$.}
\label{fig01}
\end{figure*}

\begin{figure}[!htbp]
\centering
\includegraphics[bb= 55 0 685 535, clip, width=1.0 \columnwidth]{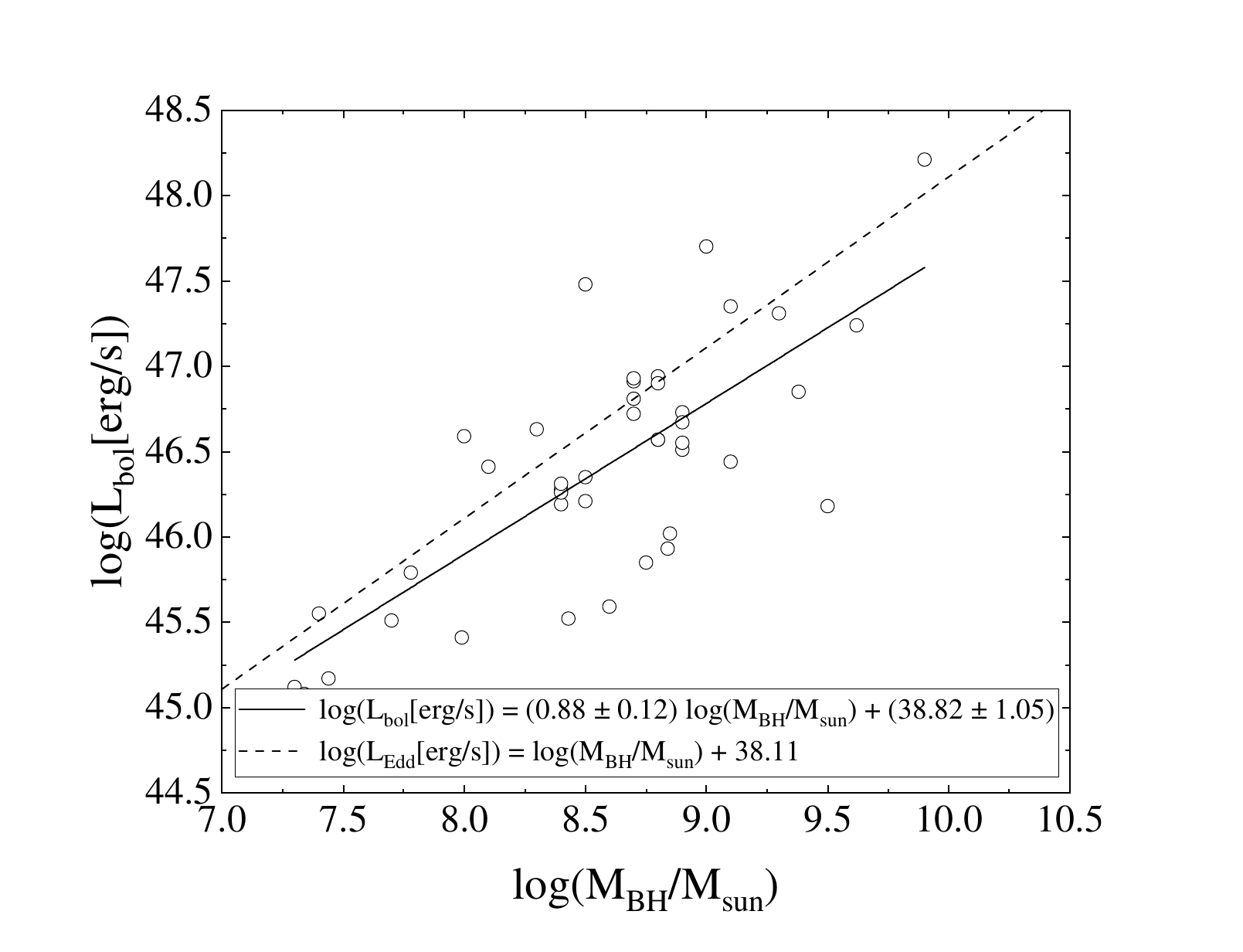}
\includegraphics[bb= 55 10 685 535, clip, width=1.0 \columnwidth]{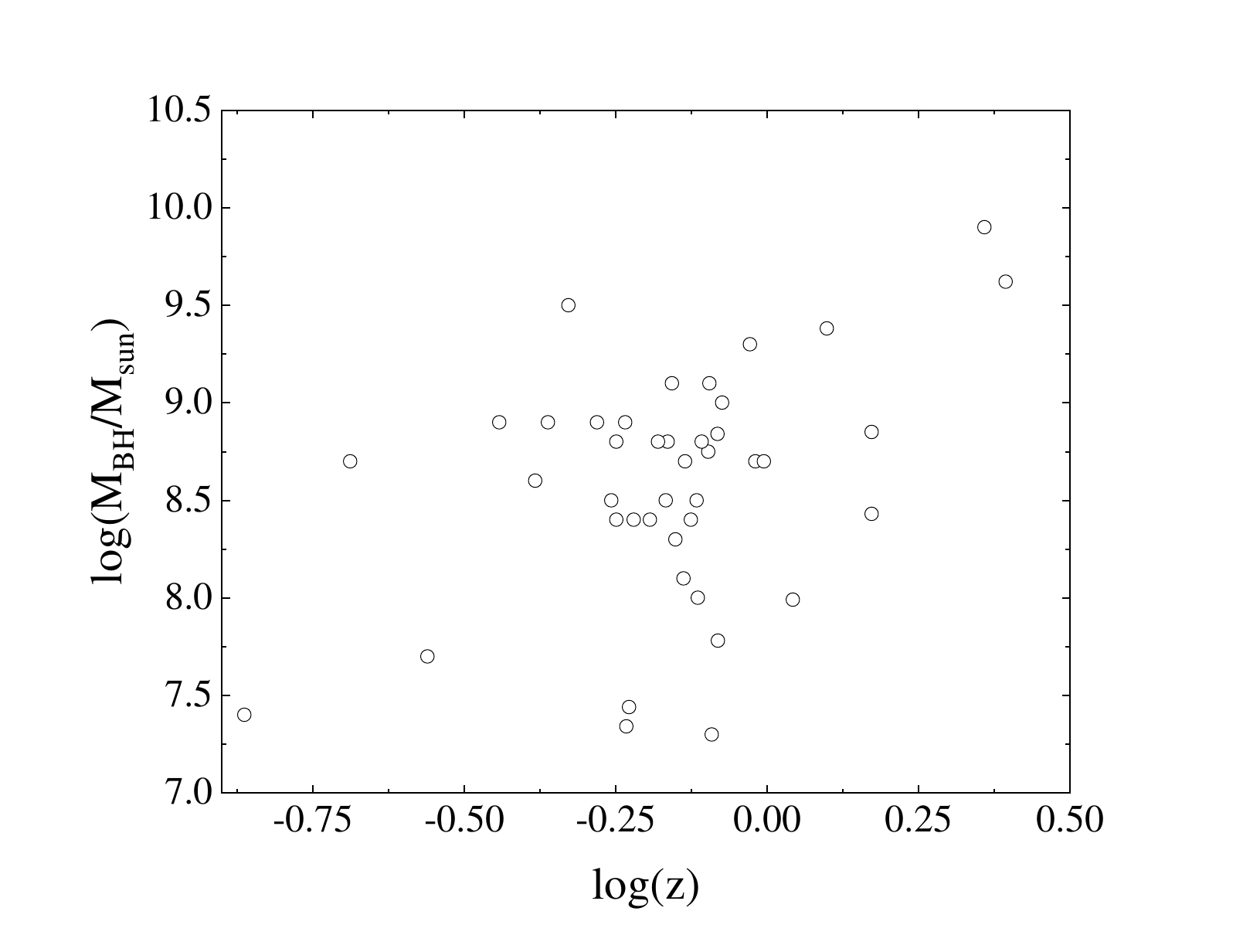}
\includegraphics[bb= 55 10 685 535, clip, width=1.0 \columnwidth]{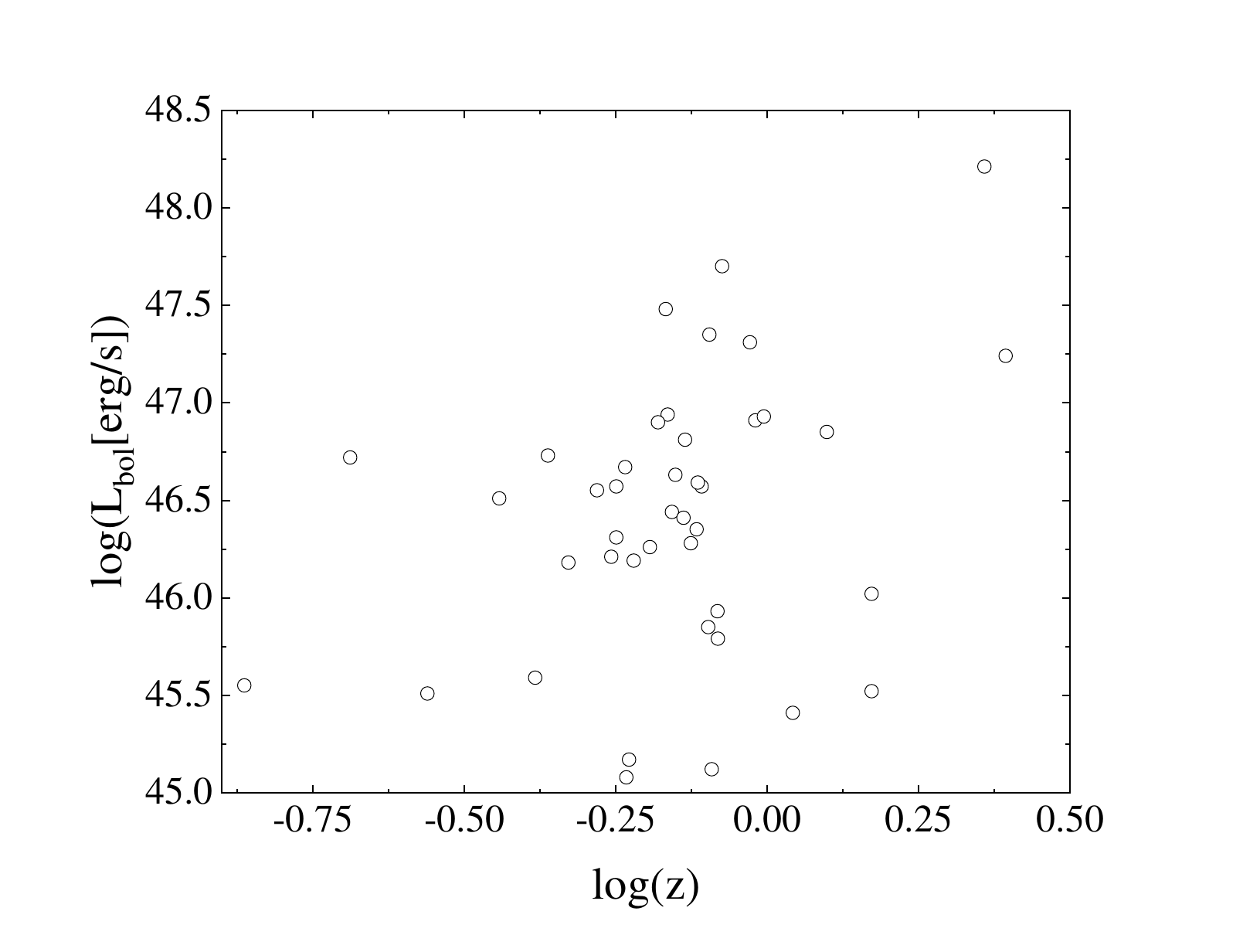}
\caption{Dependencies of the physical parameters of objects from the initial set on each other.}
\label{fig02}
\end{figure}

\section{Analysis of the initial data}

We selected a set of 42 red quasars with known physical parameters from literature \citep{glikman17,glikman24,lamassa16,lamassa17,urrutia12,kim15,kim18}. First 5 columns of Table \ref{tab1} show: object name, redshift, bolometric luminosity, Eddington ratio and SMBH mass.

Then we carried out statistical analysis of our set. Fig. \ref{fig01} shows the distributions of the parameters of our objects.

It can be seen that the SMBHs mass $M_{\rm BH}$ distribution is quite close to log-normal with a slight drop in the region of high masses and with a maximum in the region $8.5 < \log(M_{\rm BH}/M_\odot) < 9.0$, which is slightly higher than the similar value for all AGNs, which is usually taken to be equal to $\sim 8$.

The distribution of the bolometric luminosity $L_{\rm bol}$ is also close to log-normal with a drop in the region of high luminosities and with a maximum in the range $46.5 < \log(L_{\rm bol}[erg/s]) < 47.0$.

The distribution of the Eddington ratio $l_{\rm E}$ appears log-normal, but with a less pronounced peak in the region $-0.50 < \log(l_{\rm E})< -0.25$. It can be noted that the luminosity of most objects is below Eddington luminosity, but there are several objects with quite high luminosity, up to $\sim 3$ Eddington luminosities.

The cosmological redshift $z$ distribution looks quite typical for such objects. The increase at small $z$ is associated with an increase in the number of objects with distance, and the decrease at large $z$ is due to the fact that at large distances objects are much more difficult to observe.

Fig. \ref{fig02} shows the dependencies of the physical parameters of our objects on each other.

One can see that the bolometric luminosity clearly correlates with the SMBH mass (Spearman correlation coefficient is 0.65) and linear fitting gives us: $\log(L_{\rm bol}[erg/s]) = (0.88 \pm 0.12) \log(M_{\rm BH}/M_\odot) + (38.82 \pm 1.05)$, which is close to Eddington luminosity $\log(L_{\rm Edd}) = \log(M_{\rm BH}/M_\odot) + 38.11$. As for the dependencies of SMBHs mass and luminosity on redshift, it can be seen that there is no correlation between parameters (Spearmen correlation coefficients are 0.22 and 0.25 respectively). This is expected behavior for this number of objects.

\begin{figure}
\includegraphics[bb= 55 35 700 535, clip, width= 1.0\columnwidth]{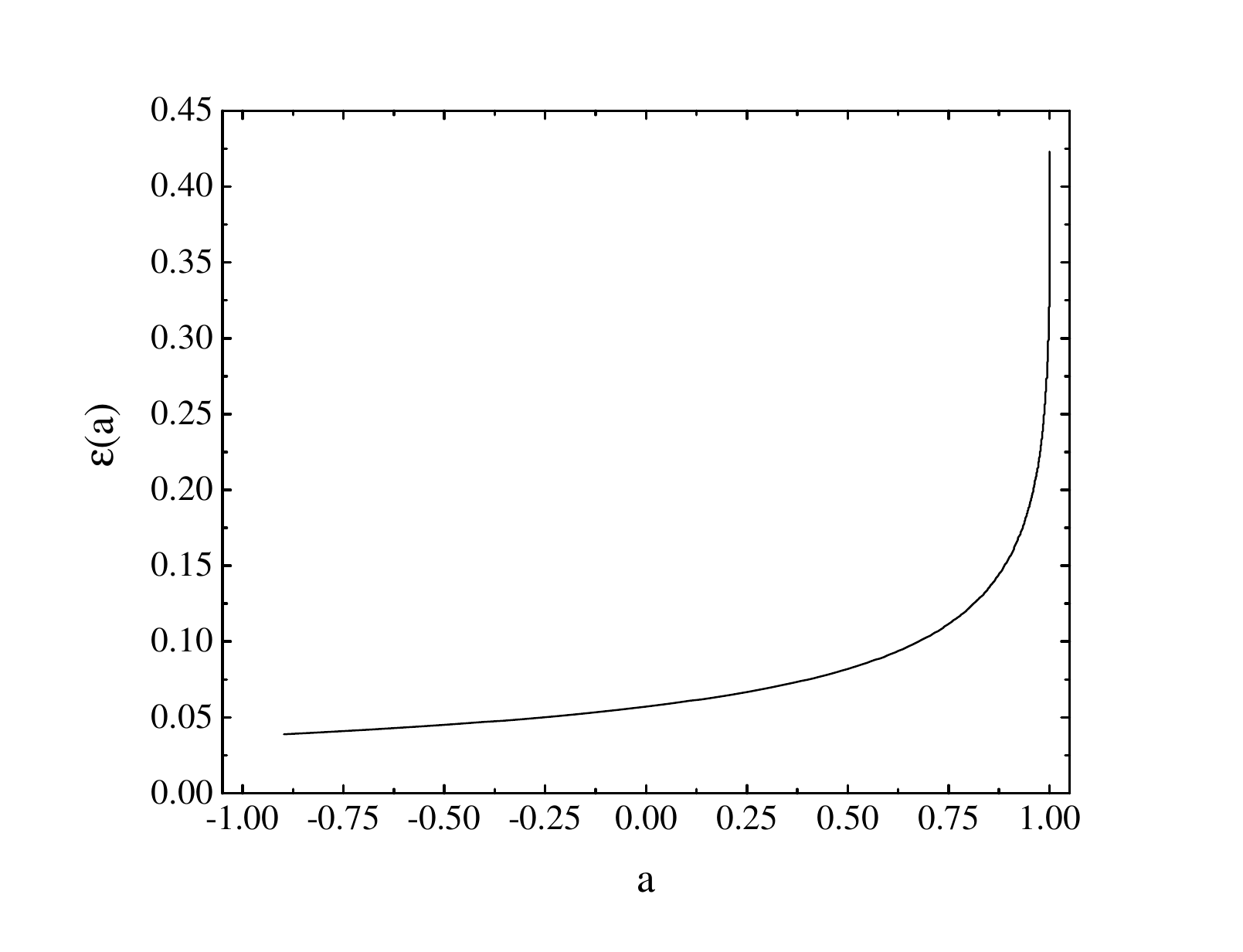}
\caption{Dependence of the radiative efficiency $\varepsilon$ on the spin $a$ {(see Eqs.(\ref{eq01},\ref{eq02}))}.}
  \label{epsilon_fig}
\end{figure}

\section{Method for estimating spin values}

One approach to estimating the spin of a black hole is by determining the radiative efficiency $\varepsilon(a)$ of its accretion disk. This efficiency depends on the value of the black hole's spin \citep{bardeen72,novikov73,krolik07,krolik07b} (see Fig. \ref{epsilon_fig}). The radiative efficiency is defined as $\varepsilon = L_{\rm bol} / (\dot{M} c^2)$, where $L_{\rm bol}$ is the AGN bolometric luminosity and $\dot{M}$ is the accretion rate. For this estimation, the value of radiative efficiency should fall within the range $0.039 < \varepsilon < 0.324$, and the spin $-1.0 < a \leq 0.998$ \citep{thorne74}. A negative value of the spin corresponds to ''retrograde'' rotation, where the SMBH and its accretion disk rotate in opposite directions.

The radiative efficiency coefficient has been related to various AGN parameters, including the SMBH mass $M_{\rm BH}$, the angle between the line of sight and the normal to the accretion disk plane $i$, and the bolometric luminosity $L_{\rm bol}$ \citep{davis11,raimundo11,du14,trakhtenbrot14,lawther17}. These relationships are based on statistical analyses of observational data and the Shakura-Sunyaev accretion disk model \citep{shakura73}. The methods proposed by \citet{davis11} and \citet{raimundo11}, as well as \citet{lawther17}, all rely on this framework. The methods by \citet{davis11} and \citet{raimundo11} are essentially equivalent, while the approach by \citet{lawther17} can be considered a variation of the earlier work by \citet{raimundo11}. Given these similarities, we have chosen to focus on three distinct models that differ from one another in meaningful ways. To facilitate comparison and analysis, we have reformulated the equations presented in the original papers to achieve greater uniformity:
\begin{enumerate}
\item \citet{du14}:\\ $\varepsilon \left( a \right) =  0.105 \left(\frac{L_\text{bol}}{10^{46}\text{erg/s}}\right) \left(\frac{L_{5100}}{10^{45}\text{erg/s}} \right)^{-1.5} M_8 \mu^{1.5}$.

\item \citet{raimundo11}:\\$\varepsilon \left( a \right) = 0.063\left( {\frac{L_\text{bol} }{10^{46}\text{erg/s}}} \right)^{0.99} \left( {\frac{L_\text{opt} }{10^{45}\text{erg/s}}} \right)^{-1.5} M_8^{0.89} \mu^{1.5}$.

\item \citet{trakhtenbrot14}:\\$\varepsilon \left( a \right) =  0.073 \left(\frac{L_\text{bol}}{10^{46}\text{erg/s}}\right) \left(\frac{\lambda L_\lambda}{10^{45}\text{erg/s}} \right)^{-1.5} \left(\frac{\lambda}{5100\text{\AA}}\right)^{-2} M_8 \mu^{1.5}\\ \lambda L_\lambda = L_{\rm opt}, \lambda = 4400\text{\AA}$.
\end{enumerate}

\noindent Here $L_{5100}$ is the luminosity at 5100\AA, $M_8 = M_\text{BH} / (10^8 M_{\odot})$ and $\mu = \cos{(i)}$.  For the model from \cite{du14}, we used the Eddington ratio $l_\text{E} = L_\text{bol} / L_\text{Edd}$, where $L_\text{Edd} = 1.3 \times 10^{38} M_\text{BH} / M_\odot$\,erg/s is the Eddington luminosity.

In the literature, we were able to find $L_{5100}$ for only 7 objects, so for the remaining objects we calculated these values using known $L_{\rm bol}$ and bolometric correction \citep{richards06}: $L_{5100} = L_{\rm bol} / 10.3$.

As for the angle $i$, its reliable determination from observational data is a rather complicated and not yet fully solved problem. Many authors often just assume that $i$ equals to some average constant value for all objects. In this work, we employed the following approach: for each object, we took the average value of the angle $i = 45^{\circ}$; if the current numerical method did not yield a physically meaningful result, then we alternately changed the angles to smaller and larger values with a step size of $5^{\circ}$ until achieving a meaningful outcome.

It should be noted that determined mass of SMBH actually depends on inclination angle $i$, since the most widely used method for determining mass uses expressions of the form \citep{decarli08}:
\begin{equation}
  M_{\rm BH} = \frac{R_{\rm BLR} V_{\rm BLR}^2}{G},
  \label{eq_M_BH}
\end{equation}

\noindent where $R_{\rm BLR}$ is the accretion disk broad-line region (BLR) scale radius, $V_{\rm BLR}$ is the typical velocity of matter in BLR and $G$ is the gravitational constant. $V_{\rm BLR}$ can be determined through observations, namely by measuring full width at half maximum (FWHM) of H$\beta$ spectral line:
\begin{equation}
  V_{\rm BLR} = f \times FWHM({\rm H}\beta).
  \label{eq_V_BLR}
\end{equation}

\noindent Here $f$ is coefficient describing the geometry of the accretion disk that, using some general assumptions (see \citet{decarli08}), can be expressed as:
\begin{equation}
  f = \left(2\sqrt{\left(\frac{H}{R}\right)^2 + \sin^2{i}}\right)^{-1},
  \label{eq_f}
\end{equation}

\noindent where $H/R$ is basically a ratio of the geometric thickness of the disk to the disk radius. In this work we assume that all objects have geometrically thin disks, so $H/R \ll 1$ and
\begin{equation}
  f \approx \frac{1}{2 \sin{i}}.
  \label{eq_f1}
\end{equation}

One of the popular methods for determining $R_{\rm BLR}$ from observations is the method from \citet{collin04}:
\begin{equation}
  R_{\rm BLR} = 32.9 \times \left(\frac{L_{5100}}{10^{44}\,{\rm erg/s}}\right)^{0.7}.
  \label{eq_R_BLR}
\end{equation}

When determining SMBH mass, it is usually assumed by default that $f = \sqrt{3}/2$ (in our case it corresponds to $i \approx 35^{\circ}$) \citep{collin04,decarli08}. Thus, when we estimated radiative efficiencies for various values of inclination angle we also for self-consistency estimated new mass values (using Eq.(\ref{eq_M_BH})), assuming that initial mass from the literature was obtained for $i \approx 35^{\circ}$.

Then the spin values were determined numerically using the method from \citet{bardeen72} {(see Fig.\ref{epsilon_fig})}:
\begin{equation}
  \varepsilon(a) = 1 - \frac{R_\text{ISCO}^{3/2} - 2 R_\text{ISCO}^{1/2} + |a|}{R_\text{ISCO}^ {3/4}\left(R_\text{ISCO}^{3/2} - 3 R_\text{ISCO}^{1/2} + 2 |a|\right)^{1/2}},
  \label{eq01}
\end{equation}

\noindent where $R_\text{ISCO}$ is the radius of the innermost stable circular orbit that depends on spin:
\begin{equation}
  \begin{array}{l}
   R_\text{ISCO}(a) = \\
   = 3 + Z_2 \pm [(3 - Z_1)(3 + Z_1 + 2 Z_2)]^{1/2},\\
   Z_1 = 1 + (1 - a^2)^{1/3}\left[(1 + a)^{1/3} + (1 - a)^{1/3}\right],\\
   Z_2 = (3 a^2 + Z_1^2)^{1/2}.
  \end{array}
  \label{eq02}
\end{equation}

\noindent Here ''-'' is used when $a \geq 0$, and ''+'' when $a < 0$.

\section{Results of our estimations}

We have estimated values of spin, mass and inclination angle for all 42 objects from our set using 3 models. The results are shown at Table \ref{tab1}.

One can see that our mass estimations are close to literature data (considering that average error in literature data is $\pm 0.1$ in logarithmic scale) for almost all objects except for F2MS~J1113+1244 and F2MS~J1434+0935, which, at the same time, have highest values of Eddington ratio $l_{\rm E} = 2.29$ and $l_{\rm E} = 3.35$ respectively. This may indicate that for objects with such high super-Eddington luminosity our assumption about geometrically thin disk is wrong {(or that the mass of these objects is determined incorrectly)}. However, we note that for super-Eddington objects with $l_{\rm E} < 2.0$ our assumption seems to work quite well.

For 6 objects (SDSS~J0036-0113, S82X~0040+0058, S82X~0118+0018, S82X~0303-0115, FBQS~J1227+3214, S82X~2328-0028) all 3 models produce a negative spin value, which may indicate the presence of ''retrograde'' rotation, {which can be evidence of a recent merger. It should be noted that some authors generally believe that the observed properties of the majority or even all red quasars are associated with recent mergers. Of this six objects, we were able to find in the literature some additional evidences of recent merger for FBQS~J1227+3214 \citep{glikman24}}.

\begin{figure}[!htbp]
\centering
\includegraphics[bb= 50 0 715 525, clip, width=1.0 \columnwidth]{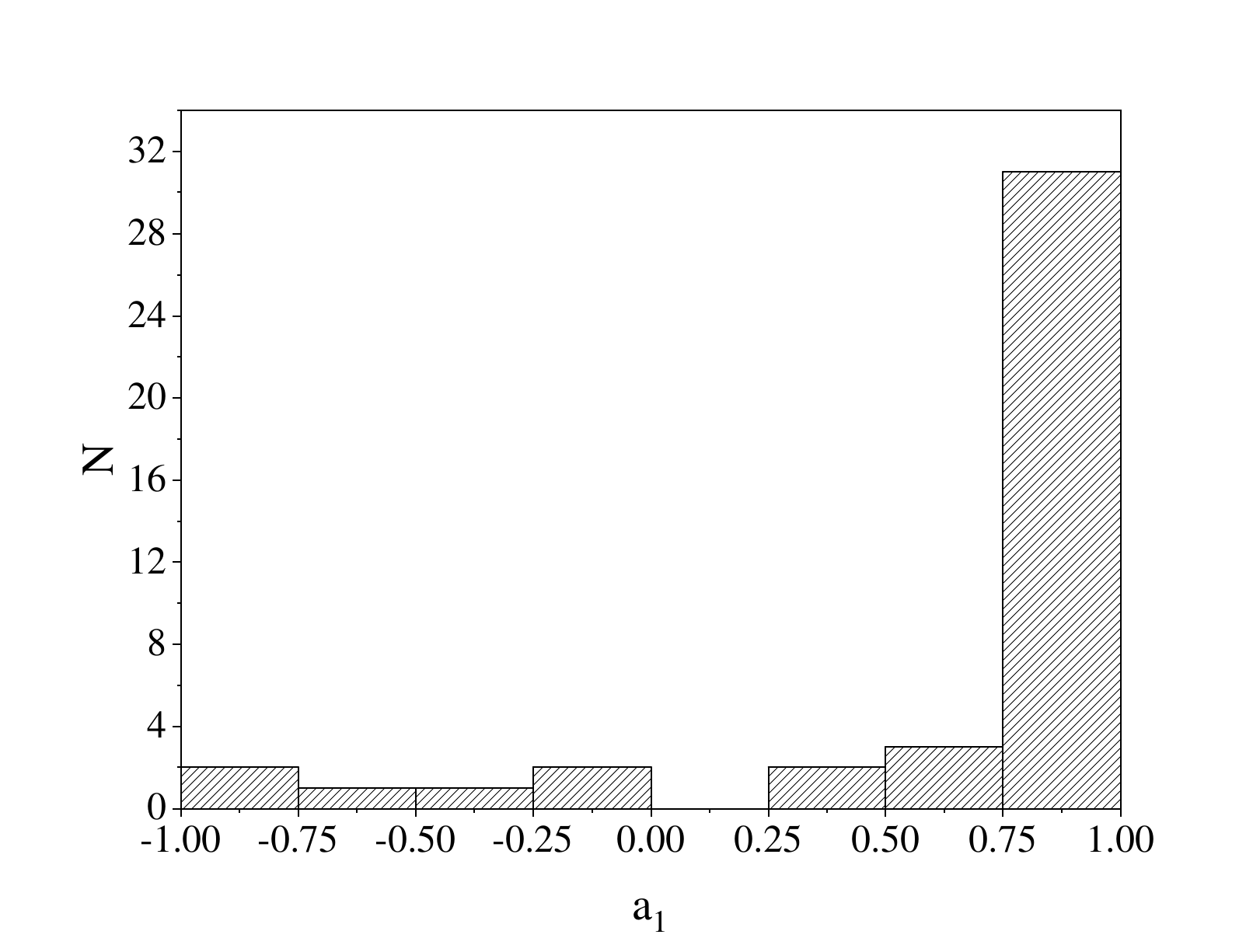}
\includegraphics[bb= 50 0 715 525, clip, width=1.0 \columnwidth]{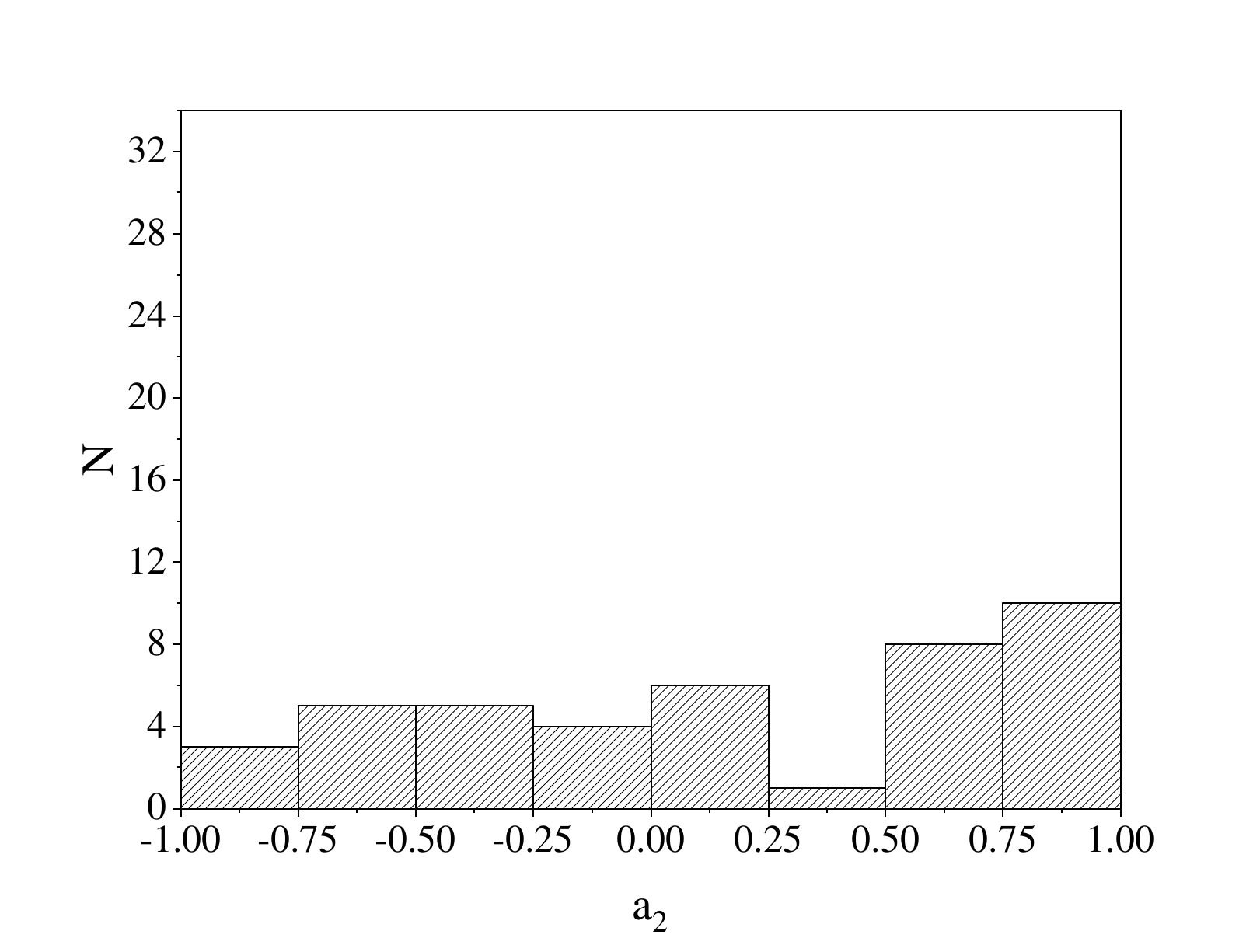}
\includegraphics[bb= 50 0 715 525, clip, width=1.0 \columnwidth]{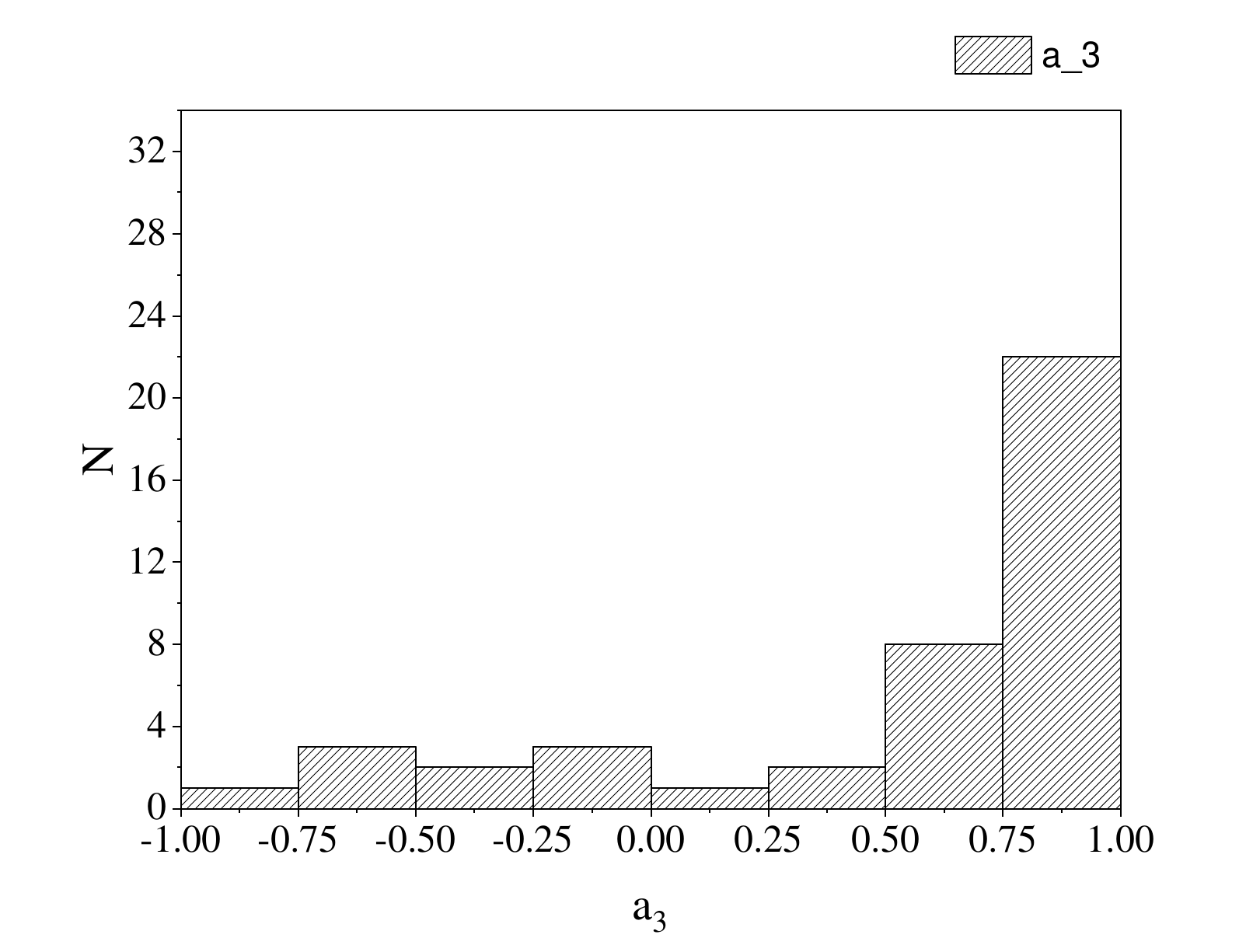}
\caption{Distributions of estimated spins for 3 models.}
\label{fig03}
\end{figure}

Fig. \ref{fig03} shows distributions of estimated spin values for 3 models. It can be seen that first and, to some extent, third models demonstrate similar distributions, which are quite characteristic of AGNs with high $z$ values \citep{trakhtenbrot14} and Seyfert galaxies \citep{afanasiev18,piotrovich22,piotrovich23}. Second model demonstrates a rather strange and atypical distribution which may indicate that this model is not suitable for determining spins of this type of objects.

\begin{figure}[!htbp]
\centering
\includegraphics[bb= 50 0 715 530, clip, width=1.0 \columnwidth]{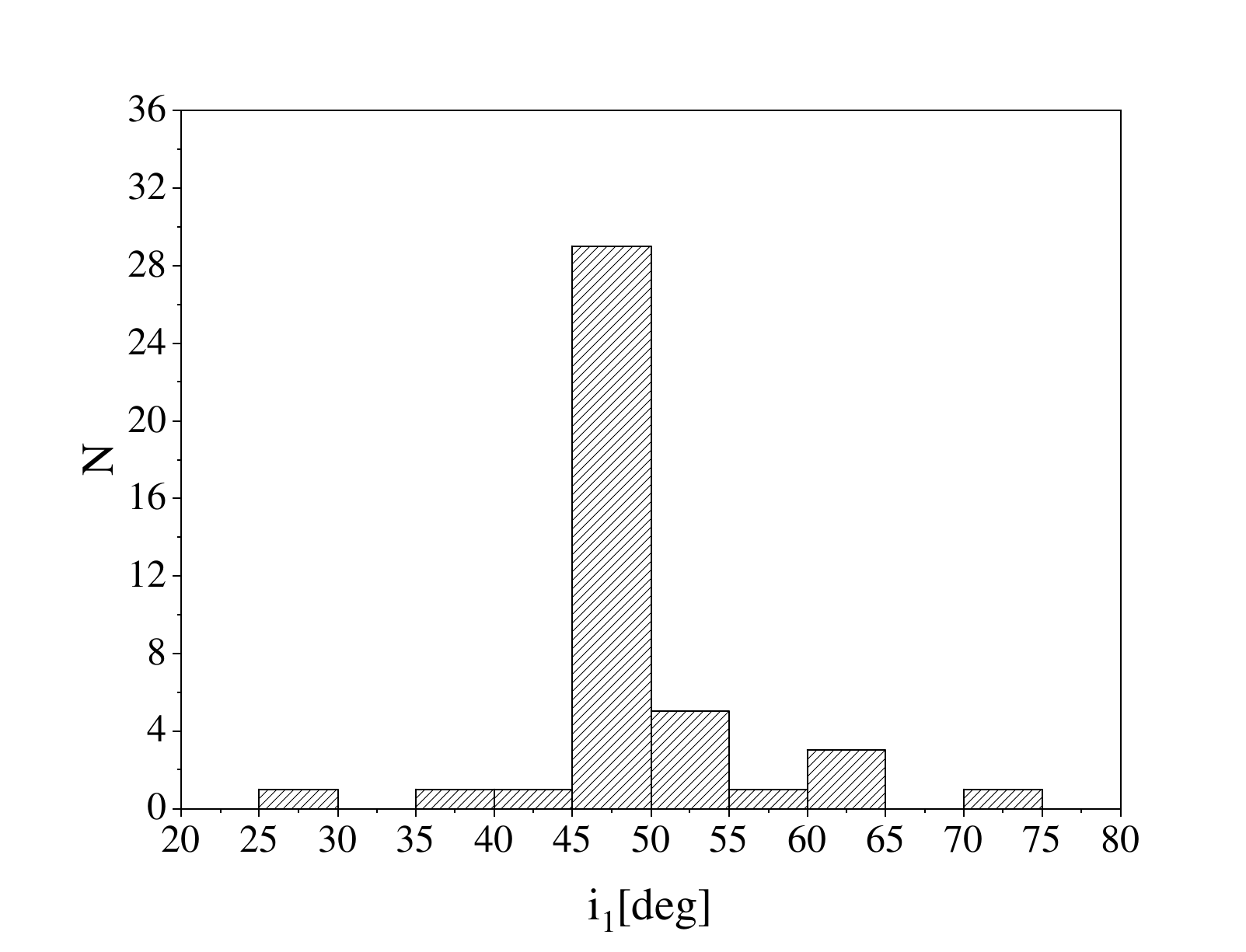}
\includegraphics[bb= 50 0 715 530, clip, width=1.0 \columnwidth]{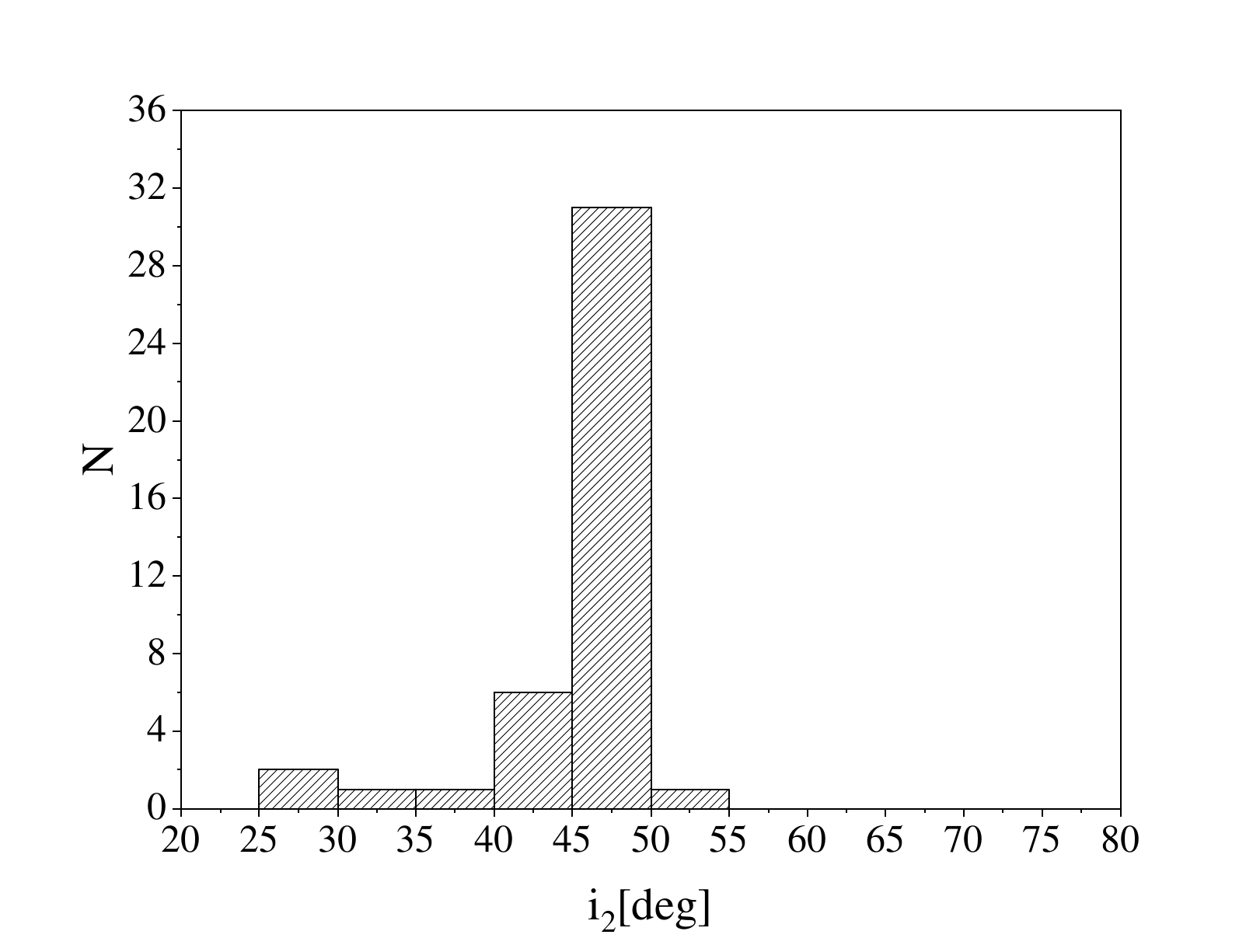}
\includegraphics[bb= 50 0 715 530, clip, width=1.0 \columnwidth]{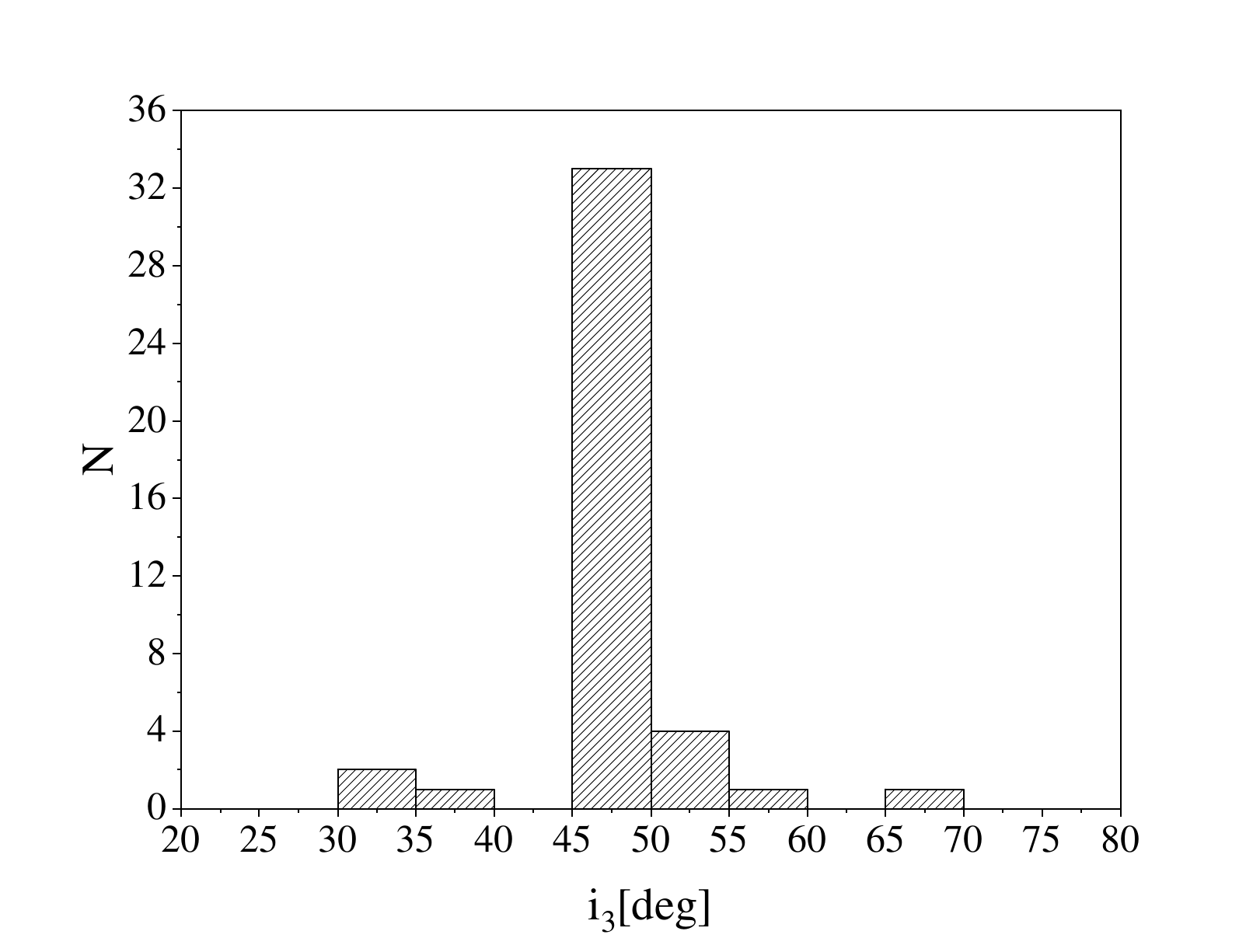}
\caption{Distributions of estimated inclination angles for 3 models.}
\label{fig04}
\end{figure}

Fig. \ref{fig04} demonstrates distribution of estimated inclination angles for 3 models. Pronounced peak near $45^{\circ}$ is associated with the calculation method we used. Our method for estimating inclination angle, of course, is not completely accurate, however it is significantly more accurate than the method that simply uses one average constant angle value for all objects.

\begin{figure}[!htbp]
\centering
\includegraphics[bb= 50 0 715 525, clip, width=1.0 \columnwidth]{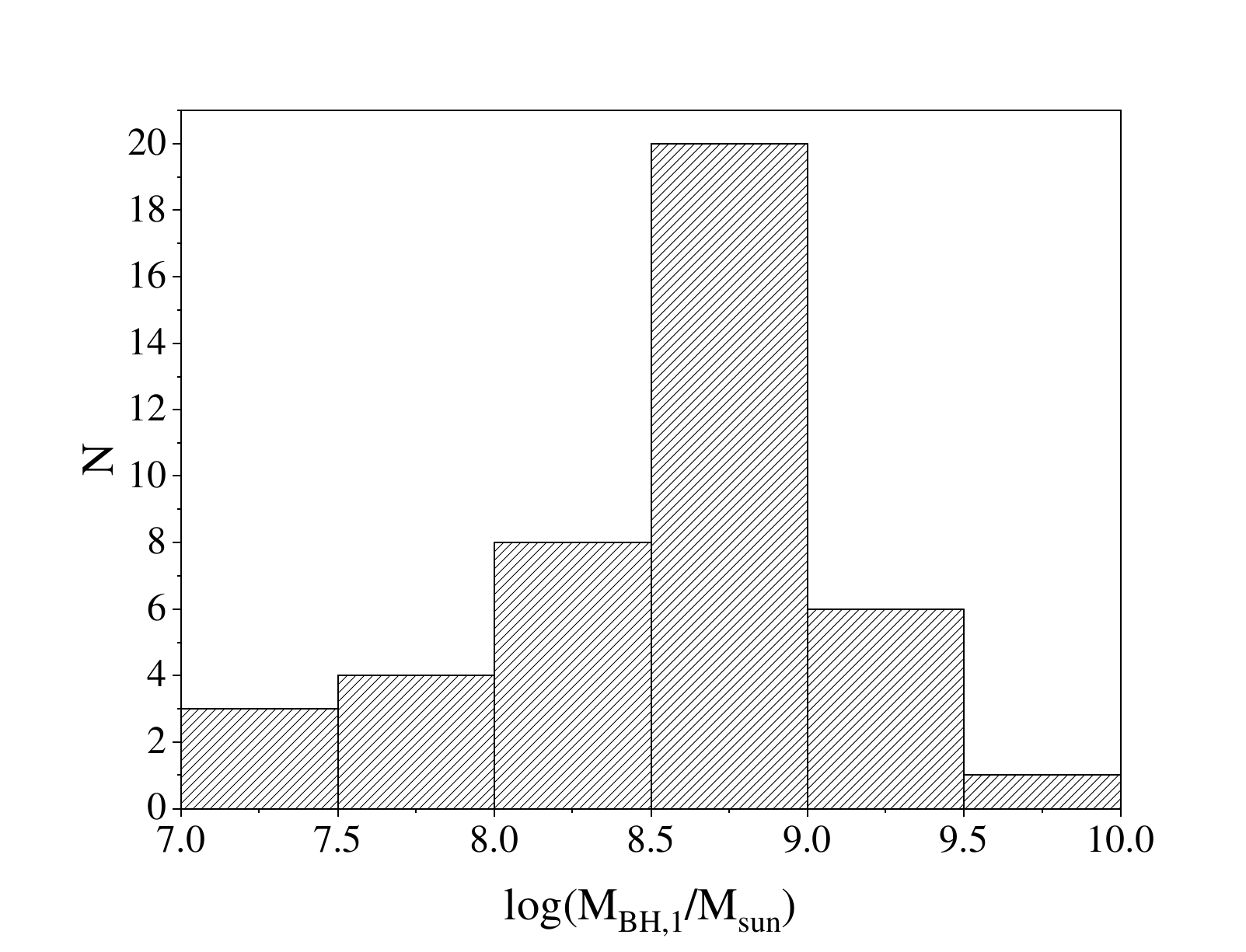}
\includegraphics[bb= 50 0 715 525, clip, width=1.0 \columnwidth]{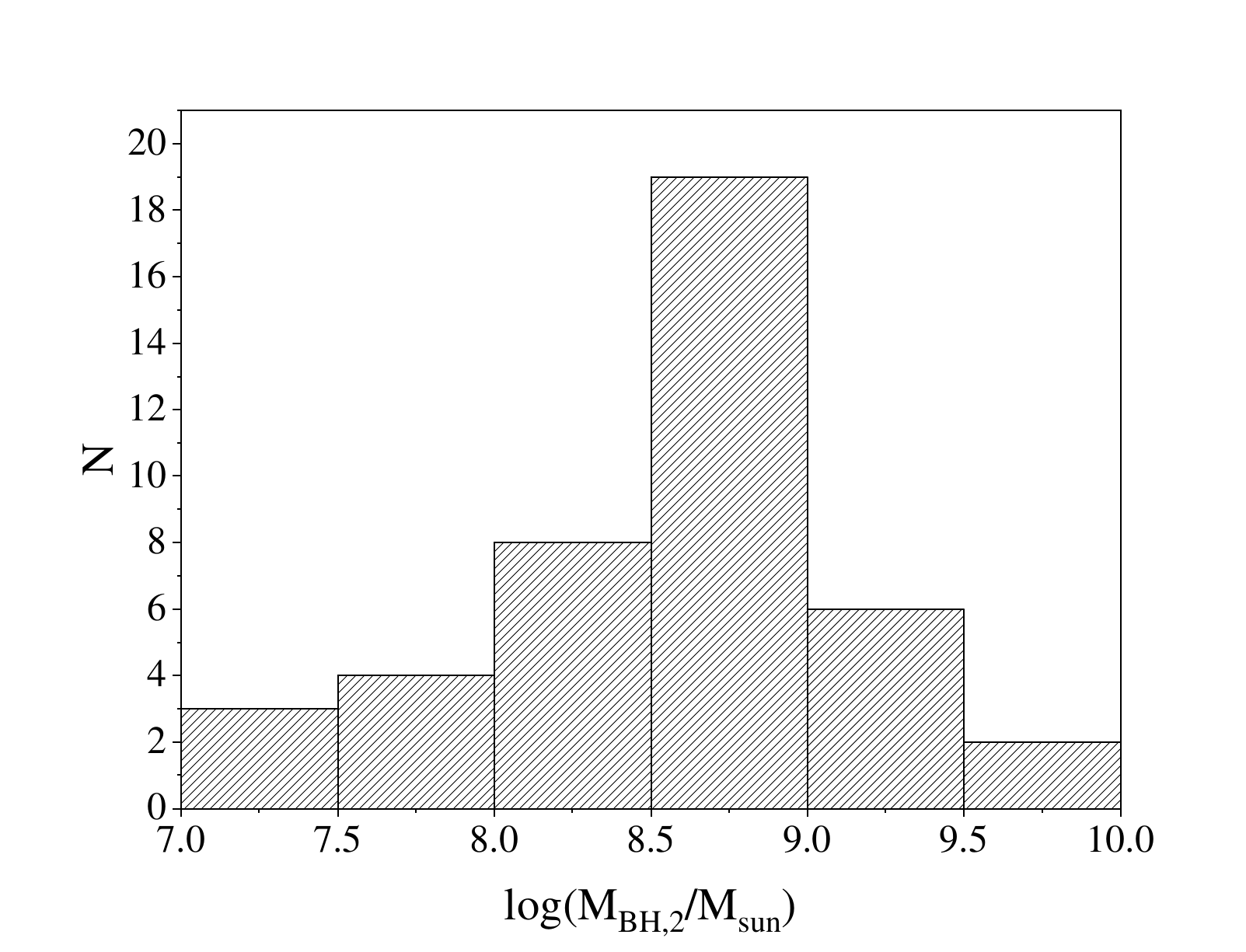}
\includegraphics[bb= 50 0 715 525, clip, width=1.0 \columnwidth]{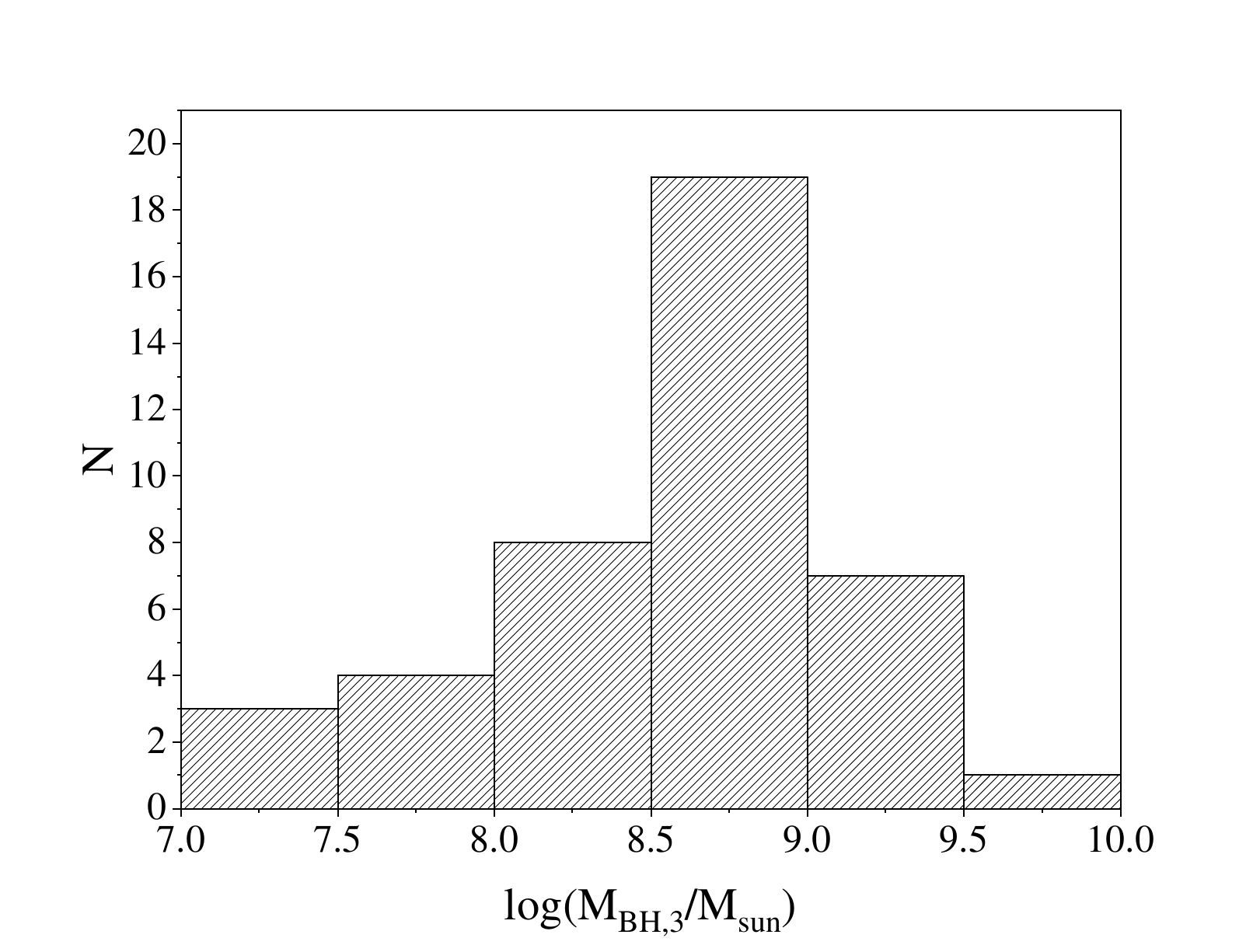}
\caption{Distributions of estimated SMBH masses for 3 models.}
\label{fig05}
\end{figure}

Fig. \ref{fig05} shows distributions of estimated SMBH masses for 3 models. All distributions look similar to distribution of initial masses from literature, what is the expected result.

\begin{figure}[!htbp]
\centering
\includegraphics[bb= 50 15 685 525, clip, width=1.0 \columnwidth]{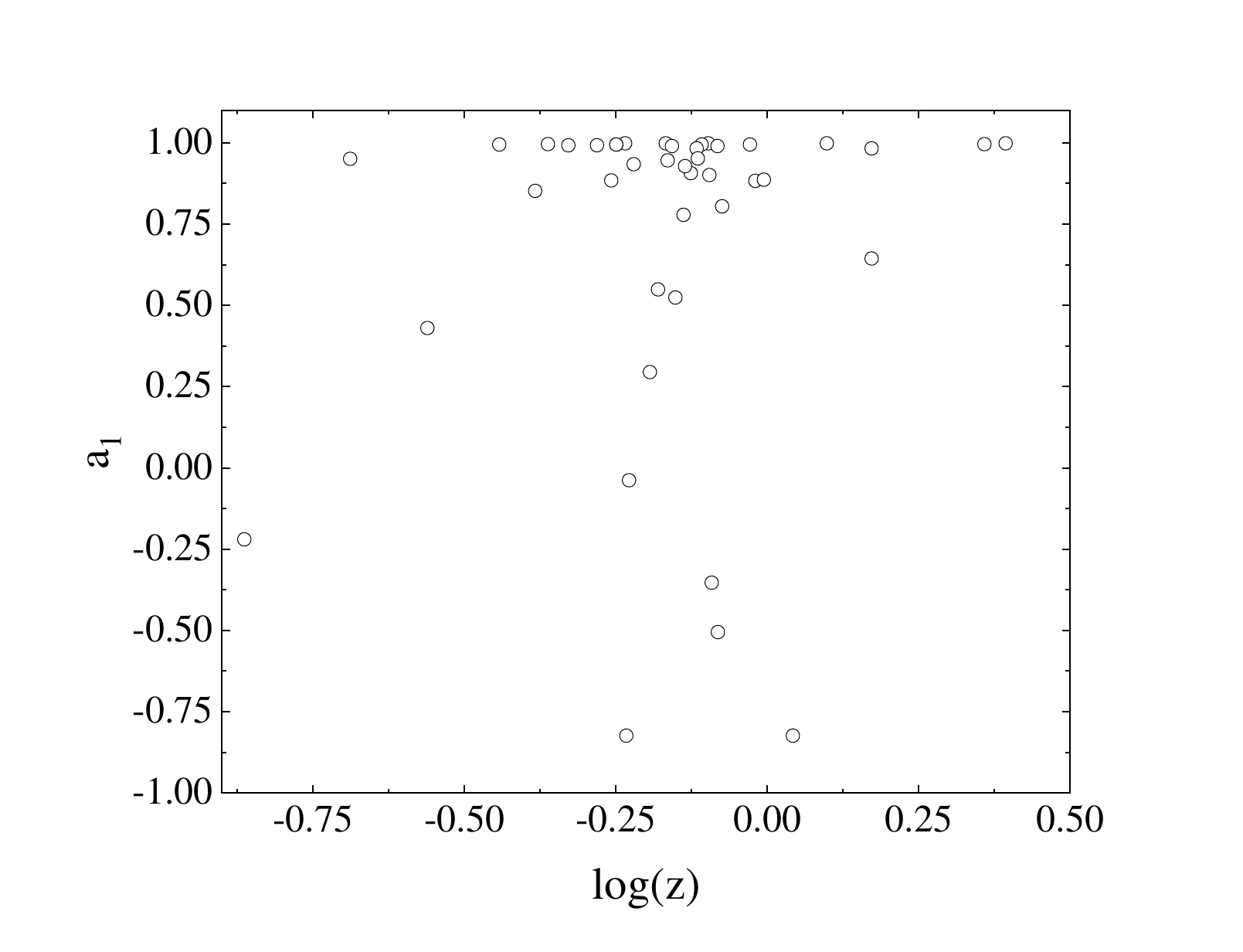}
\includegraphics[bb= 50 15 685 525, clip, width=1.0 \columnwidth]{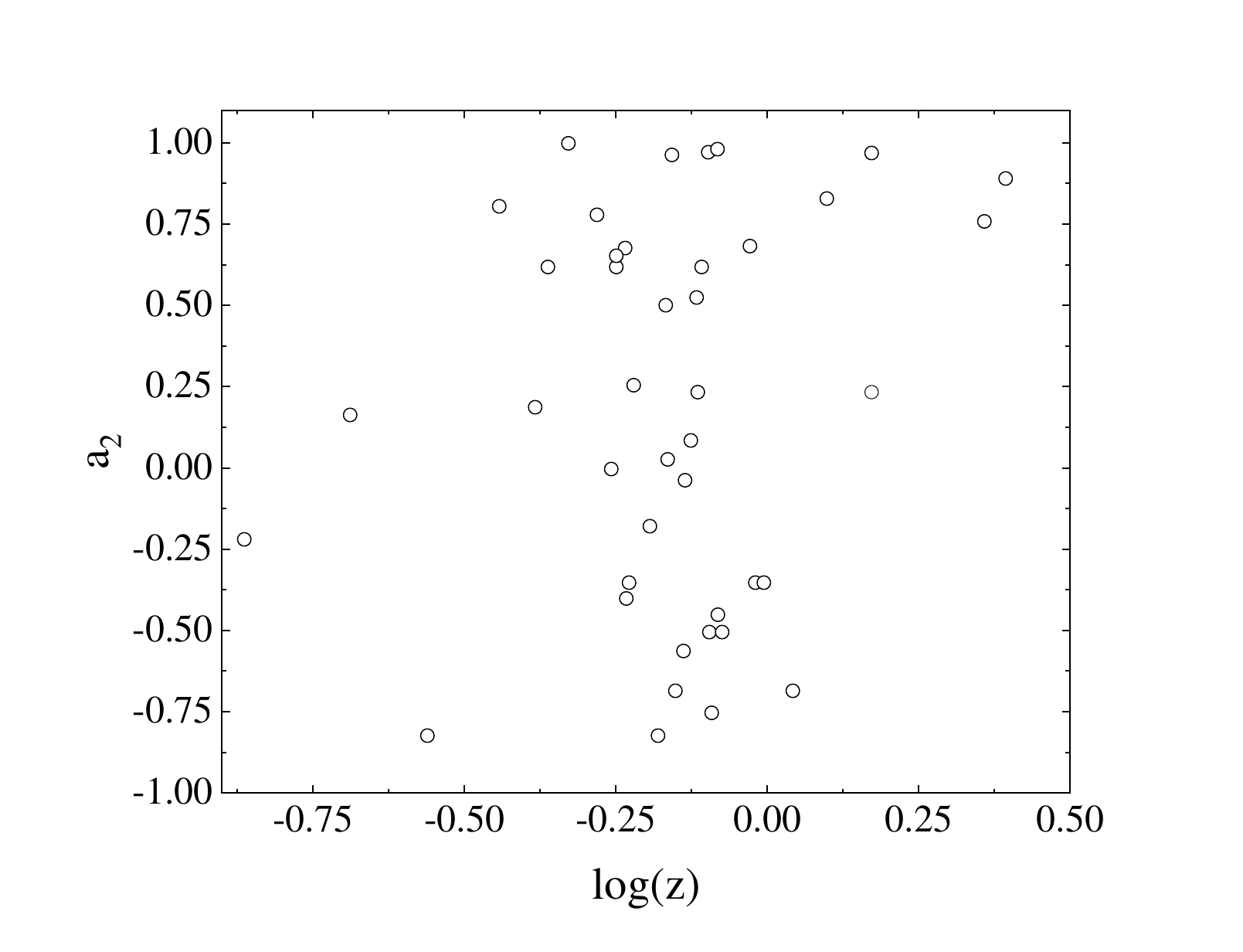}
\includegraphics[bb= 50 15 685 525, clip, width=1.0 \columnwidth]{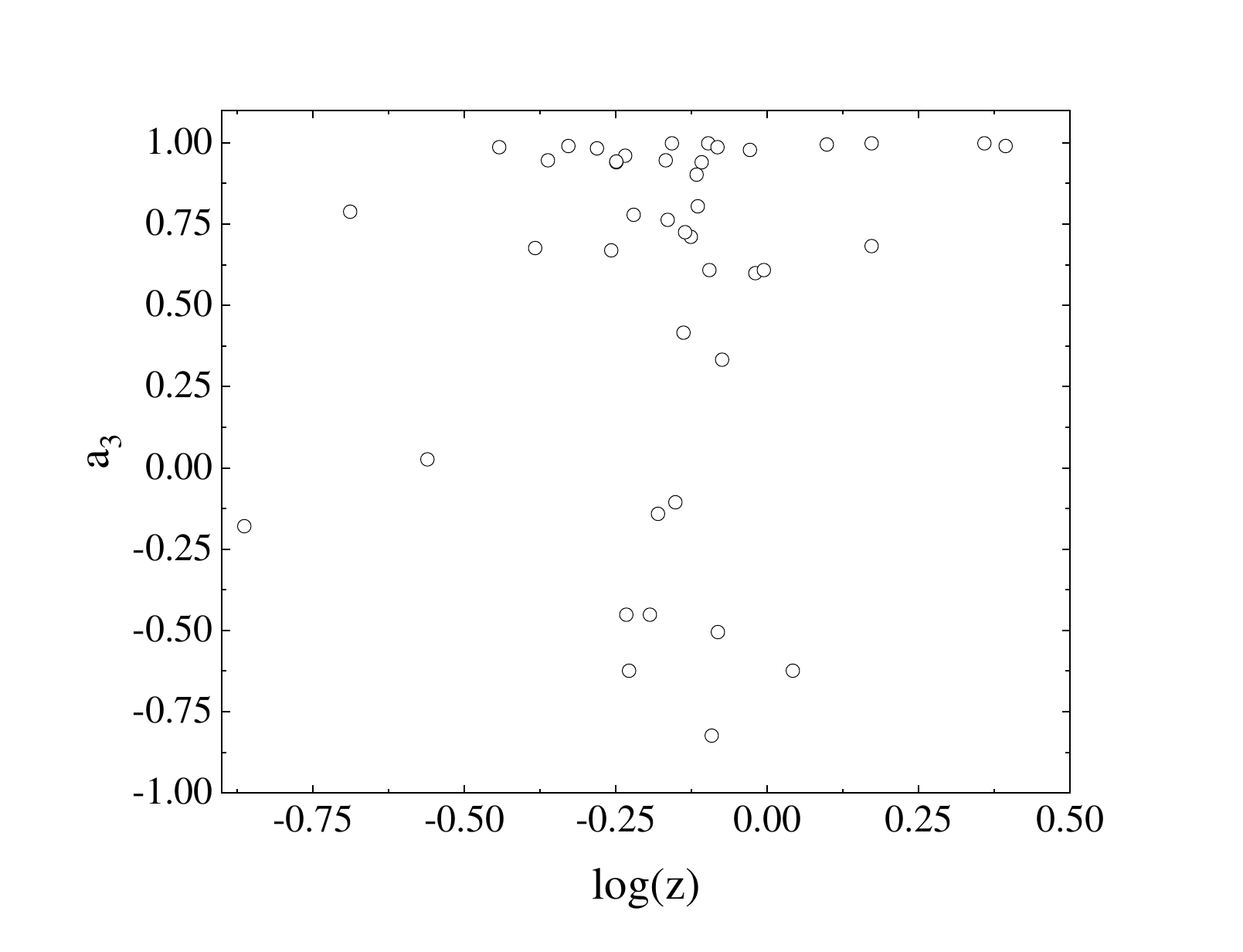}
\caption{Dependencies of the spin on the redshift for 3 models.}
\label{fig06}
\end{figure}

Fig. \ref{fig06} displays dependencies of the spin on the redshift for 3 models. There is no detectable correlation between parameters for all models (Spearman correlation coefficients are 0.06, 0.05 and 0.1 respectively). This is expected result due to the small number of objects.

\begin{figure}[!htbp]
\centering
\includegraphics[bb= 50 0 685 525, clip, width=1.0 \columnwidth]{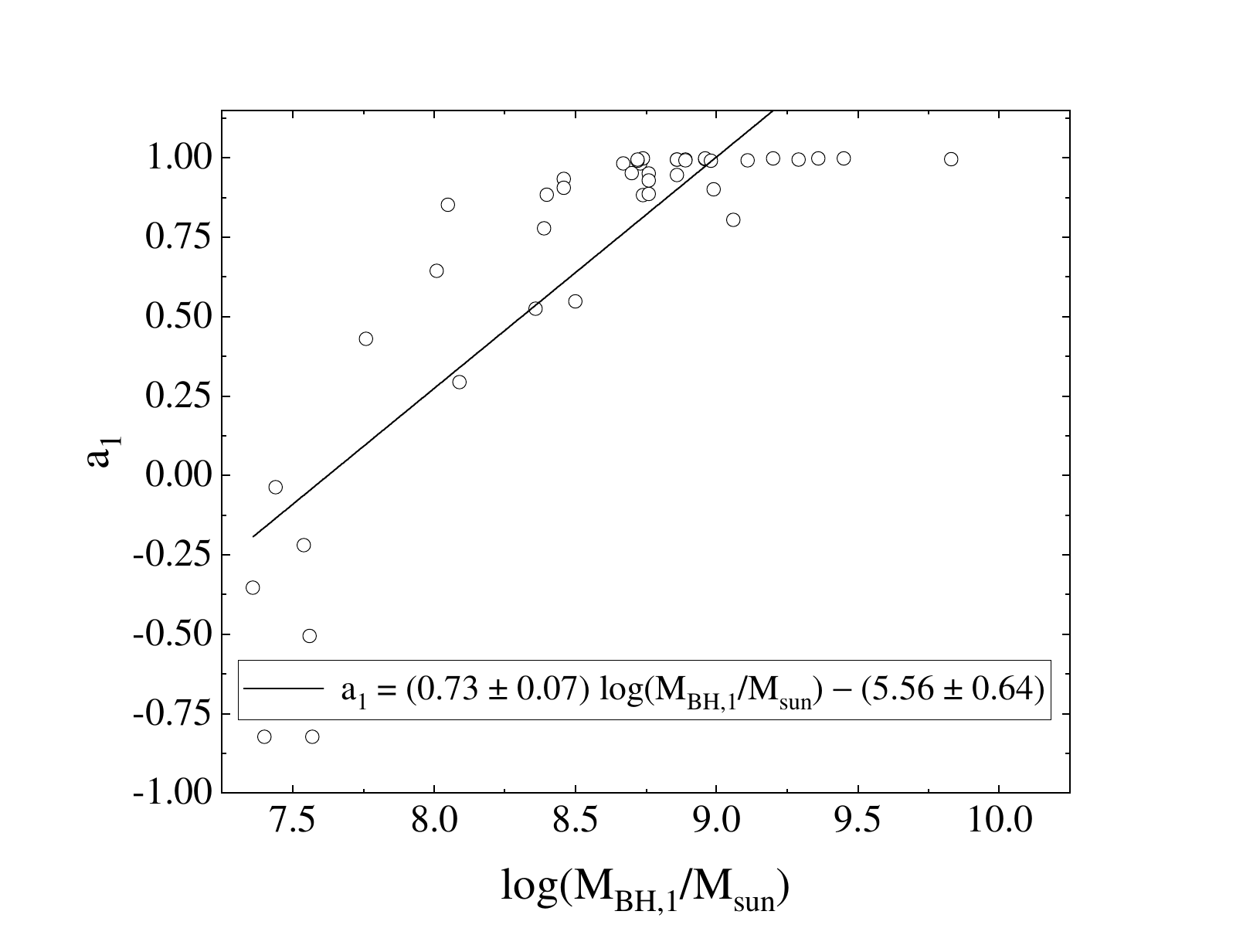}
\includegraphics[bb= 50 0 685 525, clip, width=1.0 \columnwidth]{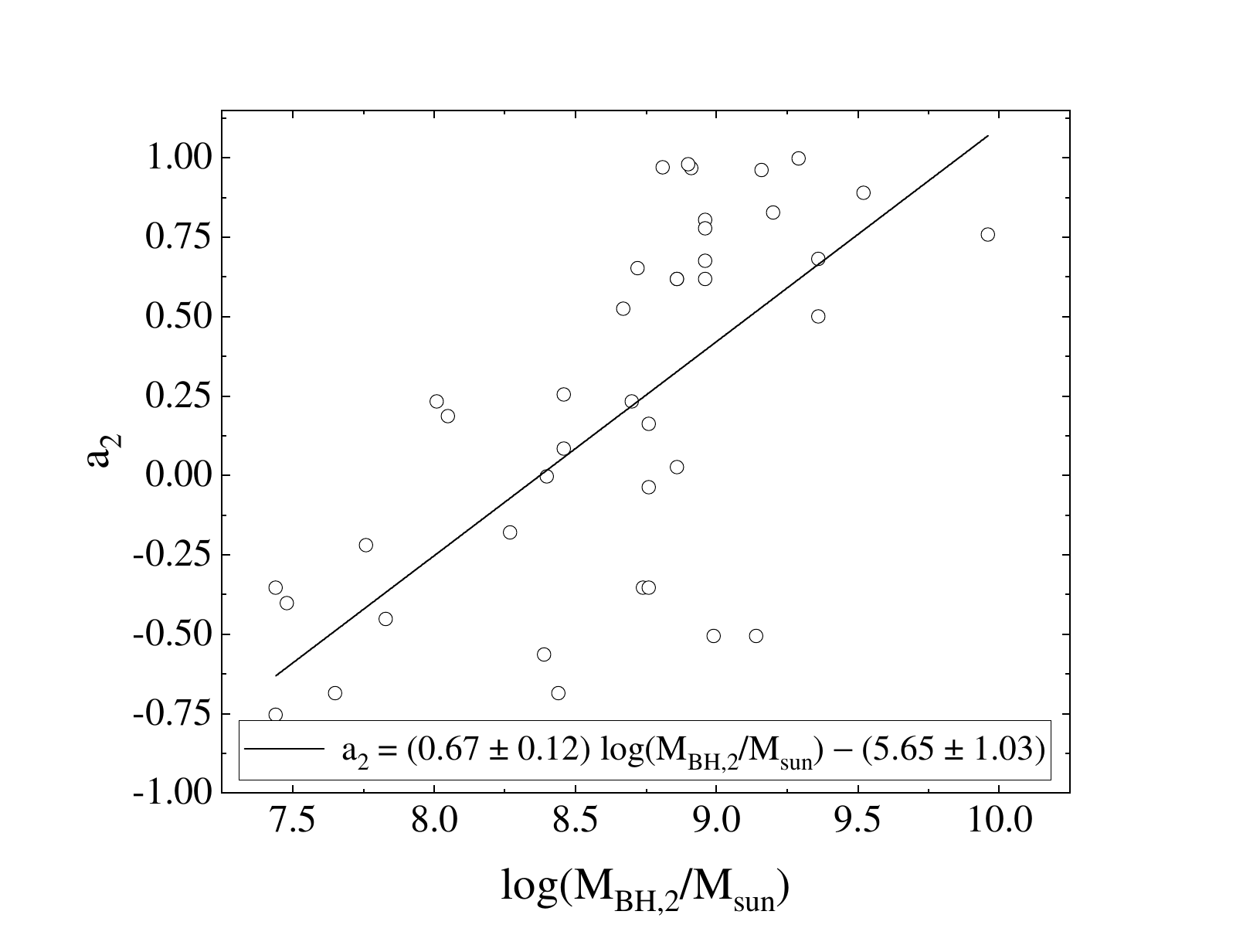}
\includegraphics[bb= 50 0 685 525, clip, width=1.0 \columnwidth]{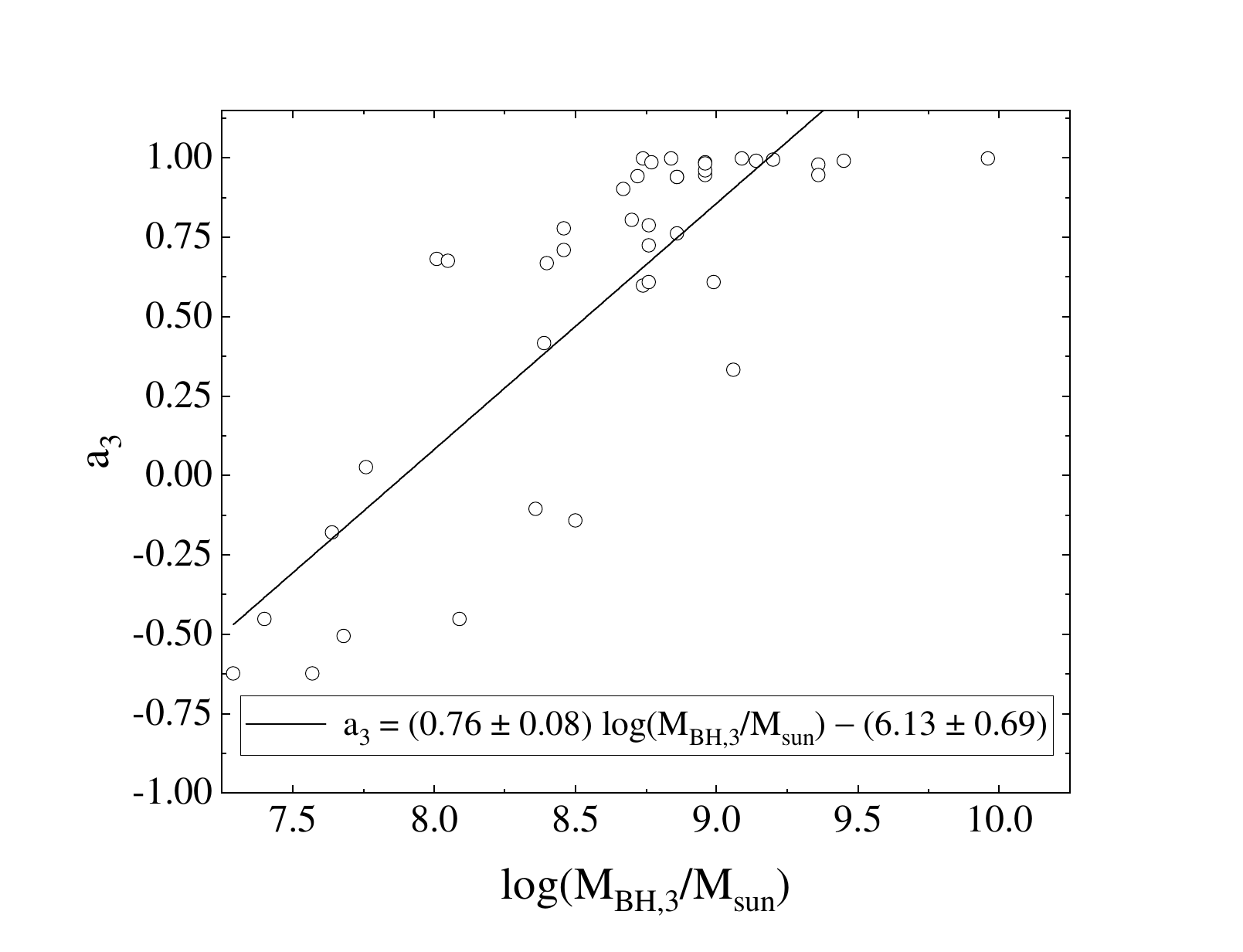}
\caption{Dependencies of the spin on the SMBH mass for 3 models.}
\label{fig07}
\end{figure}

Fig. \ref{fig07} shows dependencies of the spin on the SMBH mass for 3 models. The correlation between the parameters is prominent (Spearman correlation coefficients are 0.82, 0.67 and 0.79 respectively). The linear fitting gives us:
\begin{equation}
  \begin{aligned}
    a_1 = (0.73 \pm 0.07) \log(M_{\rm BH,1}/M_\odot) - (5.56 \pm 0.64),\\
    a_2 = (0.67 \pm 0.12) \log(M_{\rm BH,2}/M_\odot) - (5.65 \pm 1.03),\\
    a_3 = (0.76 \pm 0.08) \log(M_{\rm BH,3}/M_\odot) - (6.13 \pm 0.69).\\
  \end{aligned}
  \label{eq_a_M_BH}
\end{equation}

\noindent In our previous work \citep{piotrovich22} we obtained similar ratio for Seyfert galaxies, but with a slope value of $\sim 0.3-0.45$. And for Narrow Line Seyfert galaxies (NLS1) a slope was $\sim 1.25$ \citep{piotrovich23}. The degree of the dependence of the spin on SMBH mass shows how quickly spin grows with increasing mass and, accordingly, speaks about the mechanism of mass growth. Rapid spin growth occurs due to the disk accretion, and slow growth occurs due to the chaotic accretion (for example, SMBH mergers), during which spin can even decrease. Thus, red quasars are in the middle between Seyferts and NLS1, which may indicate, for example, that the red quasars have both types of accretion and contain both Seyferts and NLS1 type objects.

\begin{figure}[!htbp]
\centering
\includegraphics[bb= 50 0 685 525, clip, width=1.0 \columnwidth]{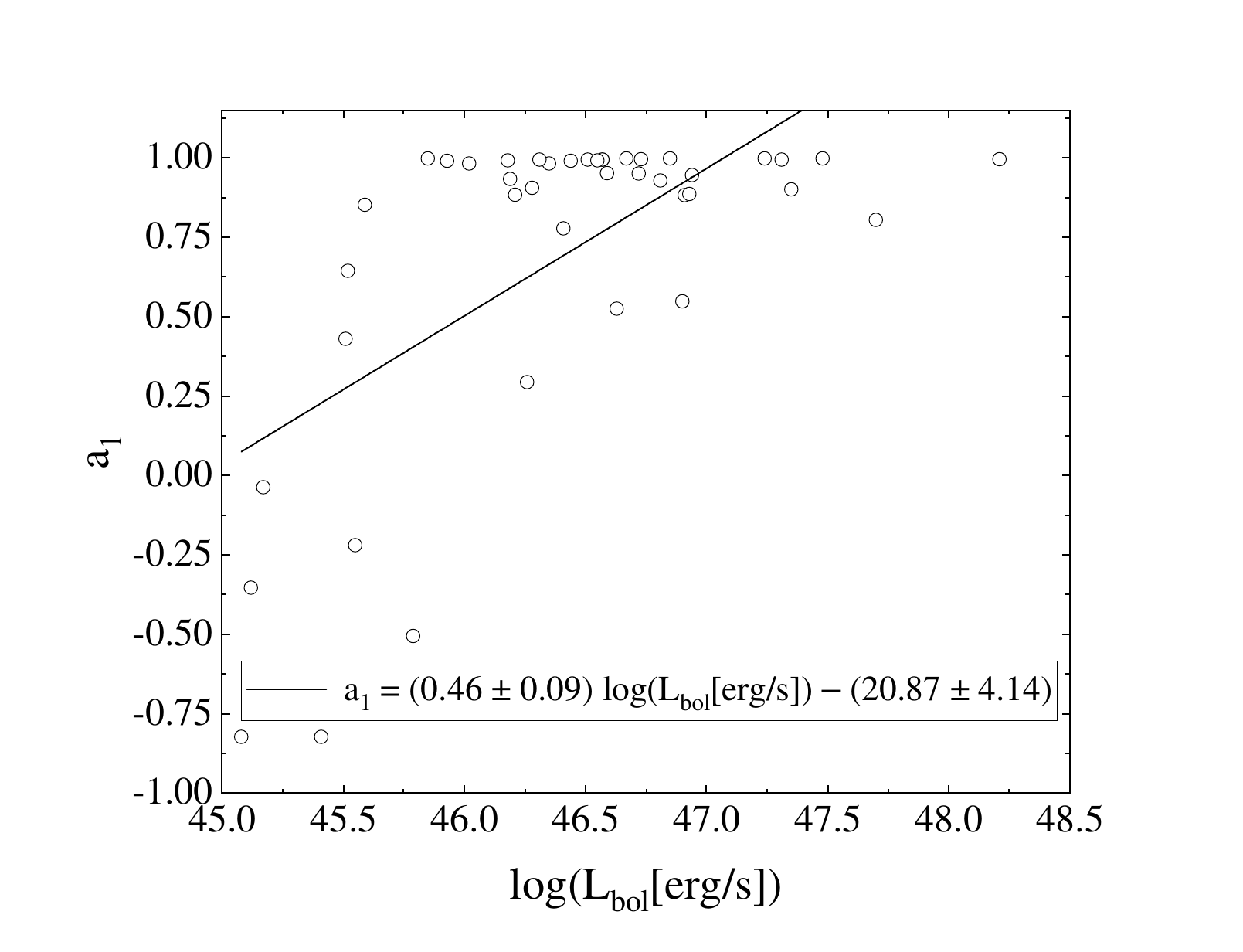}
\includegraphics[bb= 50 0 685 525, clip, width=1.0 \columnwidth]{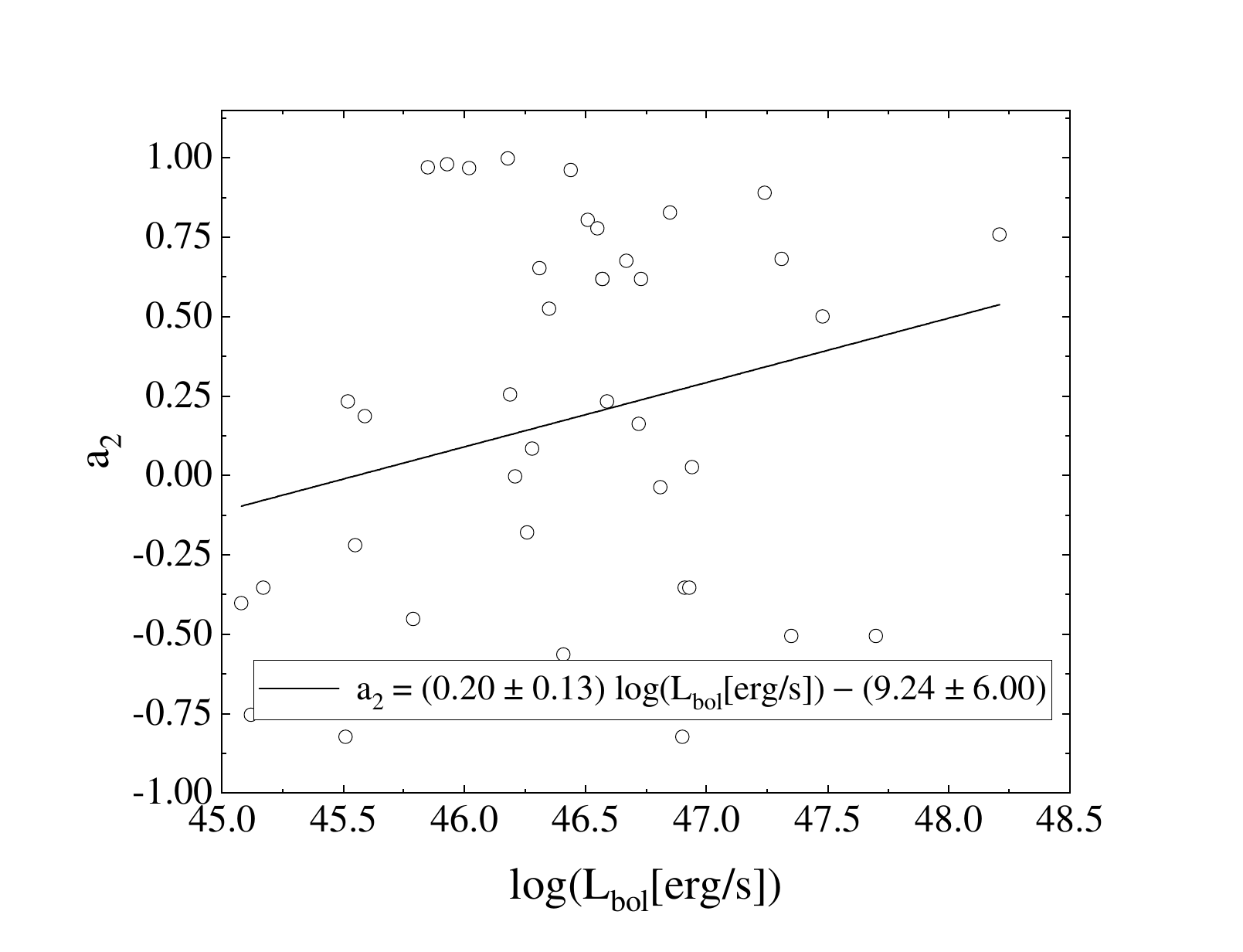}
\includegraphics[bb= 50 0 685 525, clip, width=1.0 \columnwidth]{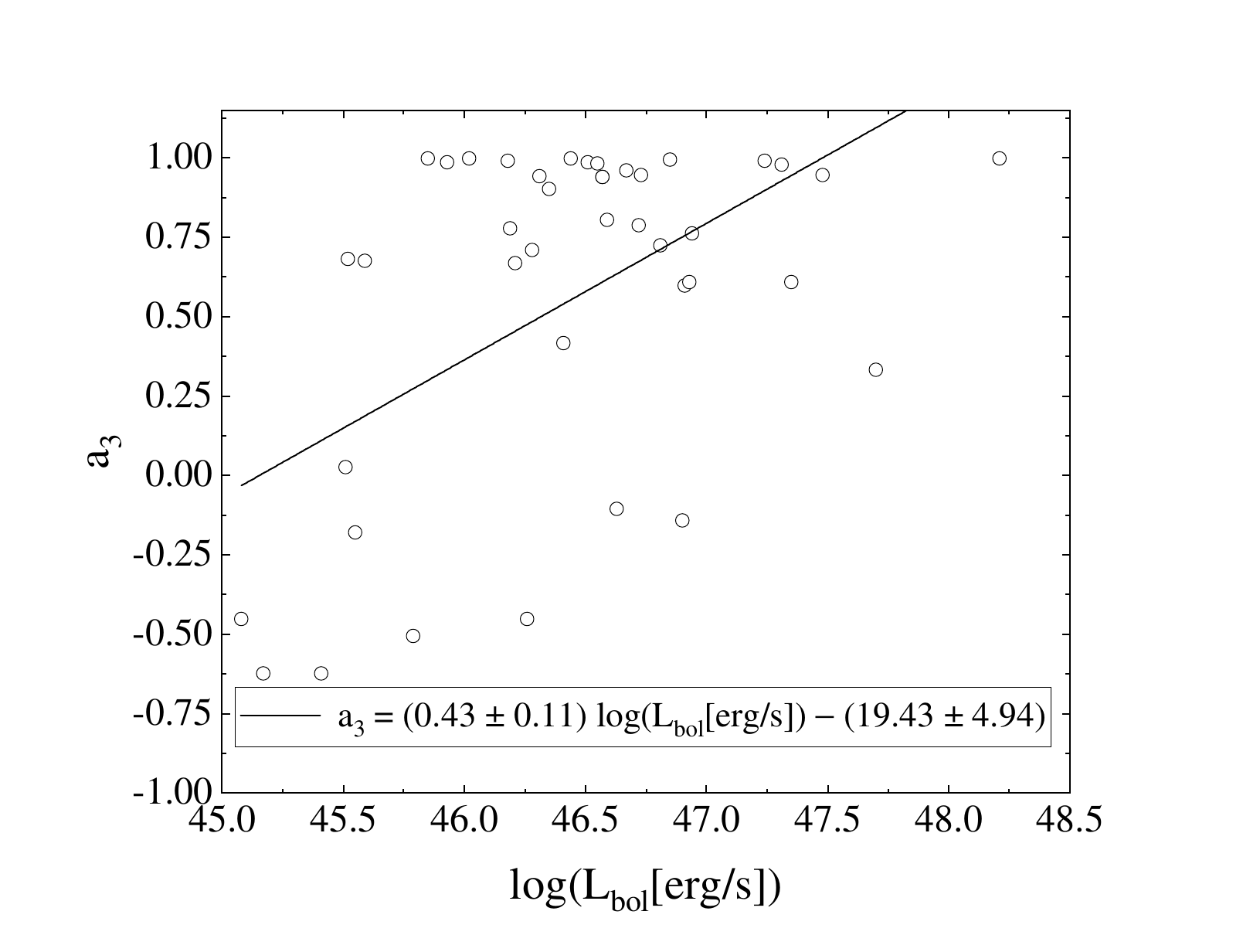}
\caption{Dependencies of the spin on the bolometric luminosity for 3 models.}
\label{fig08}
\end{figure}

Fig. \ref{fig08} demonstrates dependencies of the spin value on the bolometric luminosity for 3 models. For first model there is moderate correlation between parameters (Spearman coef. is 0.51). For second model correlation is very weak (Spearman coef. is 0.13). And for third model correlation is weak-to-moderate (Spearman coef. is 0.35). Linear fitting gives us:
\begin{equation}
  \begin{aligned}
    &a_1 = (0.46 \pm 0.09) \log(L_{\rm bol}[{\rm erg/s}]) - (20.87 \pm 4.14),\\
    &a_2 = (0.20 \pm 0.13) \log(L_{\rm bol}[{\rm erg/s}]) - (9.24 \pm 6.00),\\
    &a_3 = (0.43 \pm 0.11) \log(L_{\rm bol}[{\rm erg/s}]) - (19.43 \pm 4.94).\\
  \end{aligned}
  \label{eq_a_L_bol}
\end{equation}

\noindent Note that fittings of second and third models can only be considered as a manifestation of a general trend. Slope value for first model $0.46 \pm 0.09$ is close to similar value for NLS1 \citep{piotrovich23} $0.54 \pm 0.05$, from which it can be assumed that the population of red quasars may contain NLS1 type galaxies.

\section{Conclusions}

We estimated values of spin, mass and inclination angle for sample of 42 red quasars using 3 models \citep{du14,raimundo11,trakhtenbrot14} assuming that all objects have geometrically thin accretion disks.

Our estimations show that for 2 objects: F2MS~J1113+1244 and F2MS~J1434+0935 with highest Eddington ratios ($l_{\rm E} = 2.29$ and $l_{\rm E} = 3.35$ respectively) our approach doesn't seem to work very well, which may indicate that this two objects have geometrically thick disk {(or that the mass of these objects is determined incorrectly)}.

For six objects (SDSS~J0036-0113, S82X~0040+0058, S82X~0118+0018, S82X~0303-0115, FBQS~J1227+3214, S82X~2328-0028) all 3 models produce a negative spin value, which may indicate the presence of ''retrograde'' rotation where the SMBH and its accretion disk rotate in opposite directions, {which can be evidence of a recent merger. We were able to find in the literature some additional evidences of recent merger for object  FBQS~J1227+3214 \citep{glikman24}.}

Statistical analysis of estimated spin values and it's comparison with similar data on Seyfert galaxies (type 1, 2 and intermediate types) and Narrow Line Seyfert galaxies (NLS1) shows that red quasar population may contain Seyfert and NLS1 type galaxies with both disk and chaotic accretion types.

In general, we can say that of the three models used in this work the first model \citep{du14} seems to be best suited for objects of this type.

\section*{Acknowledgments}

This research was supported by the state order of the Central Astronomical Observatory at Pulkovo, the planned research topic ''MAGION'' - Physics and evolution of stars and active galactic nuclei.

\bibliography{paper}

\begin{thebibliography}{}

\bibitem [\protect \citeauthoryear {%
{Afanasiev}%
, {Gnedin}%
, {Piotrovich}%
, {Natsvlishvili}%
\BCBL {}\ \BBA {} {Buliga}%
}{%
{Afanasiev}%
\ \protect \BOthers {.}}{%
{\protect \APACyear {2018}}%
}]{%
afanasiev18}
\APACinsertmetastar {%
afanasiev18}%
\begin{APACrefauthors}%
{Afanasiev}, V\BPBI L.%
, {Gnedin}, Y\BPBI N.%
, {Piotrovich}, M\BPBI Y.%
, {Natsvlishvili}, T\BPBI M.%
\BCBL {}\ \BBA {} {Buliga}, S\BPBI D.%
\end{APACrefauthors}%
\unskip\
\newblock
\APACrefYearMonthDay{2018}{}{},
\newblock
\unskip
\newblock
\APACjournalVolNumPages{Astronomy Letters}{44}{}{362-369}.
\PrintBackRefs{\CurrentBib}

\bibitem [\protect \citeauthoryear {%
{Antonucci}%
}{%
{Antonucci}%
}{%
{\protect \APACyear {1993}}%
}]{%
antonucci93}
\APACinsertmetastar {%
antonucci93}%
\begin{APACrefauthors}%
{Antonucci}, R.%
\end{APACrefauthors}%
\unskip\
\newblock
\APACrefYearMonthDay{1993}{}{},
\newblock
\unskip
\newblock
\APACjournalVolNumPages{\araa}{31}{}{473-521}.
\PrintBackRefs{\CurrentBib}

\bibitem [\protect \citeauthoryear {%
{Bardeen}%
, {Press}%
\BCBL {}\ \BBA {} {Teukolsky}%
}{%
{Bardeen}%
\ \protect \BOthers {.}}{%
{\protect \APACyear {1972}}%
}]{%
bardeen72}
\APACinsertmetastar {%
bardeen72}%
\begin{APACrefauthors}%
{Bardeen}, J\BPBI M.%
, {Press}, W\BPBI H.%
\BCBL {}\ \BBA {} {Teukolsky}, S\BPBI A.%
\end{APACrefauthors}%
\unskip\
\newblock
\APACrefYearMonthDay{1972}{}{},
\newblock
\unskip
\newblock
\APACjournalVolNumPages{\apj}{178}{}{347-370}.
\PrintBackRefs{\CurrentBib}

\bibitem [\protect \citeauthoryear {%
{Blandford}%
\ \BBA {} {Payne}%
}{%
{Blandford}%
\ \BBA {} {Payne}%
}{%
{\protect \APACyear {1982}}%
}]{%
blandford82}
\APACinsertmetastar {%
blandford82}%
\begin{APACrefauthors}%
{Blandford}, R\BPBI D.%
\BCBT {}\ \BBA {} {Payne}, D\BPBI G.%
\end{APACrefauthors}%
\unskip\
\newblock
\APACrefYearMonthDay{1982}{}{},
\newblock
\unskip
\newblock
\APACjournalVolNumPages{\mnras}{199}{}{883-903}.
\PrintBackRefs{\CurrentBib}

\bibitem [\protect \citeauthoryear {%
{Blandford}%
\ \BBA {} {Znajek}%
}{%
{Blandford}%
\ \BBA {} {Znajek}%
}{%
{\protect \APACyear {1977}}%
}]{%
blandford77}
\APACinsertmetastar {%
blandford77}%
\begin{APACrefauthors}%
{Blandford}, R\BPBI D.%
\BCBT {}\ \BBA {} {Znajek}, R\BPBI L.%
\end{APACrefauthors}%
\unskip\
\newblock
\APACrefYearMonthDay{1977}{}{},
\newblock
\unskip
\newblock
\APACjournalVolNumPages{\mnras}{179}{}{433-456}.
\PrintBackRefs{\CurrentBib}

\bibitem [\protect \citeauthoryear {%
{Collin}%
\ \BBA {} {Kawaguchi}%
}{%
{Collin}%
\ \BBA {} {Kawaguchi}%
}{%
{\protect \APACyear {2004}}%
}]{%
collin04}
\APACinsertmetastar {%
collin04}%
\begin{APACrefauthors}%
{Collin}, S.%
\BCBT {}\ \BBA {} {Kawaguchi}, T.%
\end{APACrefauthors}%
\unskip\
\newblock
\APACrefYearMonthDay{2004}{}{},
\newblock
\unskip
\newblock
\APACjournalVolNumPages{\aap}{426}{}{797-808}.
\PrintBackRefs{\CurrentBib}

\bibitem [\protect \citeauthoryear {%
{Daly}%
}{%
{Daly}%
}{%
{\protect \APACyear {2011}}%
}]{%
daly11}
\APACinsertmetastar {%
daly11}%
\begin{APACrefauthors}%
{Daly}, R\BPBI A.%
\end{APACrefauthors}%
\unskip\
\newblock
\APACrefYearMonthDay{2011}{}{},
\newblock
\unskip
\newblock
\APACjournalVolNumPages{\mnras}{414}{}{1253-1262}.
\PrintBackRefs{\CurrentBib}

\bibitem [\protect \citeauthoryear {%
{Davis}%
\ \BBA {} {Laor}%
}{%
{Davis}%
\ \BBA {} {Laor}%
}{%
{\protect \APACyear {2011}}%
}]{%
davis11}
\APACinsertmetastar {%
davis11}%
\begin{APACrefauthors}%
{Davis}, S\BPBI W.%
\BCBT {}\ \BBA {} {Laor}, A.%
\end{APACrefauthors}%
\unskip\
\newblock
\APACrefYearMonthDay{2011}{}{},
\newblock
\unskip
\newblock
\APACjournalVolNumPages{\apj}{728}{}{98}.
\PrintBackRefs{\CurrentBib}

\bibitem [\protect \citeauthoryear {%
{Decarli}%
, {Labita}%
, {Treves}%
\BCBL {}\ \BBA {} {Falomo}%
}{%
{Decarli}%
\ \protect \BOthers {.}}{%
{\protect \APACyear {2008}}%
}]{%
decarli08}
\APACinsertmetastar {%
decarli08}%
\begin{APACrefauthors}%
{Decarli}, R.%
, {Labita}, M.%
, {Treves}, A.%
\BCBL {}\ \BBA {} {Falomo}, R.%
\end{APACrefauthors}%
\unskip\
\newblock
\APACrefYearMonthDay{2008}{}{},
\newblock
\unskip
\newblock
\APACjournalVolNumPages{\mnras}{387}{}{1237-1247}.
\PrintBackRefs{\CurrentBib}

\bibitem [\protect \citeauthoryear {%
{Du}%
\ \protect \BOthers {.}}{%
{Du}%
\ \protect \BOthers {.}}{%
{\protect \APACyear {2014}}%
}]{%
du14}
\APACinsertmetastar {%
du14}%
\begin{APACrefauthors}%
{Du}, P.%
, {Hu}, C.%
, {Lu}, K\BHBI X.%
\ et al.\end{APACrefauthors}%
\unskip\
\newblock
\APACrefYearMonthDay{2014}{}{},
\newblock
\unskip
\newblock
\APACjournalVolNumPages{\apj}{782}{}{45}.
\PrintBackRefs{\CurrentBib}

\bibitem [\protect \citeauthoryear {%
{Garofalo}%
, {Evans}%
\BCBL {}\ \BBA {} {Sambruna}%
}{%
{Garofalo}%
\ \protect \BOthers {.}}{%
{\protect \APACyear {2010}}%
}]{%
garofalo10}
\APACinsertmetastar {%
garofalo10}%
\begin{APACrefauthors}%
{Garofalo}, D.%
, {Evans}, D\BPBI A.%
\BCBL {}\ \BBA {} {Sambruna}, R\BPBI M.%
\end{APACrefauthors}%
\unskip\
\newblock
\APACrefYearMonthDay{2010}{}{},
\newblock
\unskip
\newblock
\APACjournalVolNumPages{\mnras}{406}{}{975-986}.
\PrintBackRefs{\CurrentBib}

\bibitem [\protect \citeauthoryear {%
{Georgakakis}%
\ \protect \BOthers {.}}{%
{Georgakakis}%
\ \protect \BOthers {.}}{%
{\protect \APACyear {2009}}%
}]{%
georgakakis09}
\APACinsertmetastar {%
georgakakis09}%
\begin{APACrefauthors}%
{Georgakakis}, A.%
, {Clements}, D\BPBI L.%
, {Bendo}, G.%
, {Rowan-Robinson}, M.%
, {Nandra}, K.%
\BCBL {}\ \BBA {} {Brotherton}, M\BPBI S.%
\end{APACrefauthors}%
\unskip\
\newblock
\APACrefYearMonthDay{2009}{}{},
\newblock
\unskip
\newblock
\APACjournalVolNumPages{\mnras}{394}{1}{533-546}.
\PrintBackRefs{\CurrentBib}

\bibitem [\protect \citeauthoryear {%
{Glikman}%
, {LaMassa}%
, {Piconcelli}%
, {Urry}%
\BCBL {}\ \BBA {} {Lacy}%
}{%
{Glikman}%
\ \protect \BOthers {.}}{%
{\protect \APACyear {2017}}%
}]{%
glikman17}
\APACinsertmetastar {%
glikman17}%
\begin{APACrefauthors}%
{Glikman}, E.%
, {LaMassa}, S.%
, {Piconcelli}, E.%
, {Urry}, M.%
\BCBL {}\ \BBA {} {Lacy}, M.%
\end{APACrefauthors}%
\unskip\
\newblock
\APACrefYearMonthDay{2017}{}{},
\newblock
\unskip
\newblock
\APACjournalVolNumPages{\apj}{847}{2}{116}.
\PrintBackRefs{\CurrentBib}

\bibitem [\protect \citeauthoryear {%
{Glikman}%
, {LaMassa}%
, {Piconcelli}%
, {Zappacosta}%
\BCBL {}\ \BBA {} {Lacy}%
}{%
{Glikman}%
\ \protect \BOthers {.}}{%
{\protect \APACyear {2024}}%
}]{%
glikman24}
\APACinsertmetastar {%
glikman24}%
\begin{APACrefauthors}%
{Glikman}, E.%
, {LaMassa}, S.%
, {Piconcelli}, E.%
, {Zappacosta}, L.%
\BCBL {}\ \BBA {} {Lacy}, M.%
\end{APACrefauthors}%
\unskip\
\newblock
\APACrefYearMonthDay{2024}{}{},
\newblock
\unskip
\newblock
\APACjournalVolNumPages{\mnras}{528}{1}{711-725}.
\PrintBackRefs{\CurrentBib}

\bibitem [\protect \citeauthoryear {%
{Glikman}%
\ \protect \BOthers {.}}{%
{Glikman}%
\ \protect \BOthers {.}}{%
{\protect \APACyear {2012}}%
}]{%
glikman12}
\APACinsertmetastar {%
glikman12}%
\begin{APACrefauthors}%
{Glikman}, E.%
, {Urrutia}, T.%
, {Lacy}, M.%
\ et al.\end{APACrefauthors}%
\unskip\
\newblock
\APACrefYearMonthDay{2012}{}{},
\newblock
\unskip
\newblock
\APACjournalVolNumPages{\apj}{757}{1}{51}.
\PrintBackRefs{\CurrentBib}

\bibitem [\protect \citeauthoryear {%
{Kim}%
\ \BBA {} {Im}%
}{%
{Kim}%
\ \BBA {} {Im}%
}{%
{\protect \APACyear {2018}}%
}]{%
kim18}
\APACinsertmetastar {%
kim18}%
\begin{APACrefauthors}%
{Kim}, D.%
\BCBT {}\ \BBA {} {Im}, M.%
\end{APACrefauthors}%
\unskip\
\newblock
\APACrefYearMonthDay{2018}{}{},
\newblock
\unskip
\newblock
\APACjournalVolNumPages{\aap}{610}{}{A31}.
\PrintBackRefs{\CurrentBib}

\bibitem [\protect \citeauthoryear {%
{Kim}%
, {Im}%
, {Glikman}%
, {Woo}%
\BCBL {}\ \BBA {} {Urrutia}%
}{%
{Kim}%
\ \protect \BOthers {.}}{%
{\protect \APACyear {2015}}%
}]{%
kim15}
\APACinsertmetastar {%
kim15}%
\begin{APACrefauthors}%
{Kim}, D.%
, {Im}, M.%
, {Glikman}, E.%
, {Woo}, J\BHBI H.%
\BCBL {}\ \BBA {} {Urrutia}, T.%
\end{APACrefauthors}%
\unskip\
\newblock
\APACrefYearMonthDay{2015}{}{},
\newblock
\unskip
\newblock
\APACjournalVolNumPages{\apj}{812}{1}{66}.
\PrintBackRefs{\CurrentBib}

\bibitem [\protect \citeauthoryear {%
{Krolik}%
}{%
{Krolik}%
}{%
{\protect \APACyear {2007}}%
}]{%
krolik07}
\APACinsertmetastar {%
krolik07}%
\begin{APACrefauthors}%
{Krolik}, J\BPBI H.%
\end{APACrefauthors}%
\unskip\
\newblock
\APACrefYearMonthDay{2007}{}{},
\newblock
{\BBOQ}\APACrefatitle {{Making black holes visible: accretion, radiation, and
  jets}} {{Making black holes visible: accretion, radiation, and jets}}.{\BBCQ}
\newblock
\BIn{} \APACrefbtitle {{2007 STScI Spring Symposium on Black Holes}} {{2007
  STScI Spring Symposium on Black Holes}}\ \BPG~309-321.
\PrintBackRefs{\CurrentBib}

\bibitem [\protect \citeauthoryear {%
{Krolik}%
, {Hawley}%
\BCBL {}\ \BBA {} {Hirose}%
}{%
{Krolik}%
\ \protect \BOthers {.}}{%
{\protect \APACyear {2007}}%
}]{%
krolik07b}
\APACinsertmetastar {%
krolik07b}%
\begin{APACrefauthors}%
{Krolik}, J\BPBI H.%
, {Hawley}, J\BPBI F.%
\BCBL {}\ \BBA {} {Hirose}, S.%
\end{APACrefauthors}%
\unskip\
\newblock
\APACrefYearMonthDay{2007}{}{},
\newblock
{\BBOQ}\APACrefatitle {{The Relationship between Accretion Disks and Jets}}
  {{The Relationship between Accretion Disks and Jets}}.{\BBCQ}
\newblock
\BIn{} \APACrefbtitle {Revista Mexicana de Astronomia y Astrofisica, vol. 27}
  {Revista Mexicana de Astronomia y Astrofisica, vol. 27}\ \BVOL~27, \BPG~1-7.
\PrintBackRefs{\CurrentBib}

\bibitem [\protect \citeauthoryear {%
{LaMassa}%
\ \protect \BOthers {.}}{%
{LaMassa}%
\ \protect \BOthers {.}}{%
{\protect \APACyear {2017}}%
}]{%
lamassa17}
\APACinsertmetastar {%
lamassa17}%
\begin{APACrefauthors}%
{LaMassa}, S\BPBI M.%
, {Glikman}, E.%
, {Brusa}, M.%
\ et al.\end{APACrefauthors}%
\unskip\
\newblock
\APACrefYearMonthDay{2017}{}{},
\newblock
\unskip
\newblock
\APACjournalVolNumPages{\apj}{847}{2}{100}.
\PrintBackRefs{\CurrentBib}

\bibitem [\protect \citeauthoryear {%
{LaMassa}%
\ \protect \BOthers {.}}{%
{LaMassa}%
\ \protect \BOthers {.}}{%
{\protect \APACyear {2016}}%
}]{%
lamassa16}
\APACinsertmetastar {%
lamassa16}%
\begin{APACrefauthors}%
{LaMassa}, S\BPBI M.%
, {Ricarte}, A.%
, {Glikman}, E.%
\ et al.\end{APACrefauthors}%
\unskip\
\newblock
\APACrefYearMonthDay{2016}{}{},
\newblock
\unskip
\newblock
\APACjournalVolNumPages{\apj}{820}{1}{70}.
\PrintBackRefs{\CurrentBib}

\bibitem [\protect \citeauthoryear {%
{Lawther}%
, {Vestergaard}%
, {Raimundo}%
\BCBL {}\ \BBA {} {Grupe}%
}{%
{Lawther}%
\ \protect \BOthers {.}}{%
{\protect \APACyear {2017}}%
}]{%
lawther17}
\APACinsertmetastar {%
lawther17}%
\begin{APACrefauthors}%
{Lawther}, D.%
, {Vestergaard}, M.%
, {Raimundo}, S.%
\BCBL {}\ \BBA {} {Grupe}, D.%
\end{APACrefauthors}%
\unskip\
\newblock
\APACrefYearMonthDay{2017}{}{},
\newblock
\unskip
\newblock
\APACjournalVolNumPages{\mnras}{467}{4}{4674-4710}.
\PrintBackRefs{\CurrentBib}

\bibitem [\protect \citeauthoryear {%
{Lyke}%
\ \protect \BOthers {.}}{%
{Lyke}%
\ \protect \BOthers {.}}{%
{\protect \APACyear {2020}}%
}]{%
lyke20}
\APACinsertmetastar {%
lyke20}%
\begin{APACrefauthors}%
{Lyke}, B\BPBI W.%
, {Higley}, A\BPBI N.%
, {McLane}, J\BPBI N.%
\ et al.\end{APACrefauthors}%
\unskip\
\newblock
\APACrefYearMonthDay{2020}{}{},
\newblock
\unskip
\newblock
\APACjournalVolNumPages{\apjs}{250}{1}{8}.
\PrintBackRefs{\CurrentBib}

\bibitem [\protect \citeauthoryear {%
{Novikov}%
\ \BBA {} {Thorne}%
}{%
{Novikov}%
\ \BBA {} {Thorne}%
}{%
{\protect \APACyear {1973}}%
}]{%
novikov73}
\APACinsertmetastar {%
novikov73}%
\begin{APACrefauthors}%
{Novikov}, I\BPBI D.%
\BCBT {}\ \BBA {} {Thorne}, K\BPBI S.%
\end{APACrefauthors}%
\unskip\
\newblock
\APACrefYearMonthDay{1973}{}{},
\newblock
{\BBOQ}\APACrefatitle {{Astrophysics of black holes.}} {{Astrophysics of black
  holes.}}{\BBCQ}
\newblock
\BIn{} C.~{Dewitt}\ \BBA {} B\BPBI S.~{Dewitt}\ (\BEDS), \APACrefbtitle {Black
  Holes (Les Astres Occlus)} {Black Holes (Les Astres Occlus)}\ \BPG~343-450.
\newblock
\APACaddressPublisher{New York}{Gordon and Breach}.
\PrintBackRefs{\CurrentBib}

\bibitem [\protect \citeauthoryear {%
M.~{Piotrovich}%
, {Buliga}%
\BCBL {}\ \BBA {} {Natsvlishvili}%
}{%
M.~{Piotrovich}%
\ \protect \BOthers {.}}{%
{\protect \APACyear {2023}}%
}]{%
piotrovich23}
\APACinsertmetastar {%
piotrovich23}%
\begin{APACrefauthors}%
{Piotrovich}, M.%
, {Buliga}, S.%
\BCBL {}\ \BBA {} {Natsvlishvili}, T.%
\end{APACrefauthors}%
\unskip\
\newblock
\APACrefYearMonthDay{2023}{}{},
\newblock
\unskip
\newblock
\APACjournalVolNumPages{Universe}{9}{4}{175}.
\PrintBackRefs{\CurrentBib}

\bibitem [\protect \citeauthoryear {%
M\BPBI Y.~{Piotrovich}%
, {Buliga}%
\BCBL {}\ \BBA {} {Natsvlishvili}%
}{%
M\BPBI Y.~{Piotrovich}%
\ \protect \BOthers {.}}{%
{\protect \APACyear {2022}}%
}]{%
piotrovich22}
\APACinsertmetastar {%
piotrovich22}%
\begin{APACrefauthors}%
{Piotrovich}, M\BPBI Y.%
, {Buliga}, S\BPBI D.%
\BCBL {}\ \BBA {} {Natsvlishvili}, T\BPBI M.%
\end{APACrefauthors}%
\unskip\
\newblock
\APACrefYearMonthDay{2022}{}{},
\newblock
\unskip
\newblock
\APACjournalVolNumPages{Astronomische Nachrichten}{343}{5}{e10020}.
\PrintBackRefs{\CurrentBib}

\bibitem [\protect \citeauthoryear {%
{Raimundo}%
, {Fabian}%
, {Vasudevan}%
, {Gandhi}%
\BCBL {}\ \BBA {} {Wu}%
}{%
{Raimundo}%
\ \protect \BOthers {.}}{%
{\protect \APACyear {2012}}%
}]{%
raimundo11}
\APACinsertmetastar {%
raimundo11}%
\begin{APACrefauthors}%
{Raimundo}, S\BPBI I.%
, {Fabian}, A\BPBI C.%
, {Vasudevan}, R\BPBI V.%
, {Gandhi}, P.%
\BCBL {}\ \BBA {} {Wu}, J.%
\end{APACrefauthors}%
\unskip\
\newblock
\APACrefYearMonthDay{2012}{}{},
\newblock
\unskip
\newblock
\APACjournalVolNumPages{\mnras}{419}{}{2529-2544}.
\PrintBackRefs{\CurrentBib}

\bibitem [\protect \citeauthoryear {%
{Richards}%
\ \protect \BOthers {.}}{%
{Richards}%
\ \protect \BOthers {.}}{%
{\protect \APACyear {2006}}%
}]{%
richards06}
\APACinsertmetastar {%
richards06}%
\begin{APACrefauthors}%
{Richards}, G\BPBI T.%
, {Lacy}, M.%
, {Storrie-Lombardi}, L\BPBI J.%
\ et al.\end{APACrefauthors}%
\unskip\
\newblock
\APACrefYearMonthDay{2006}{}{},
\newblock
\unskip
\newblock
\APACjournalVolNumPages{\apjs}{166}{}{470-497}.
\PrintBackRefs{\CurrentBib}

\bibitem [\protect \citeauthoryear {%
{Shakura}%
\ \BBA {} {Sunyaev}%
}{%
{Shakura}%
\ \BBA {} {Sunyaev}%
}{%
{\protect \APACyear {1973}}%
}]{%
shakura73}
\APACinsertmetastar {%
shakura73}%
\begin{APACrefauthors}%
{Shakura}, N\BPBI I.%
\BCBT {}\ \BBA {} {Sunyaev}, R\BPBI A.%
\end{APACrefauthors}%
\unskip\
\newblock
\APACrefYearMonthDay{1973}{}{},
\newblock
\unskip
\newblock
\APACjournalVolNumPages{\aap}{24}{}{337-355}.
\PrintBackRefs{\CurrentBib}

\bibitem [\protect \citeauthoryear {%
{Thorne}%
}{%
{Thorne}%
}{%
{\protect \APACyear {1974}}%
}]{%
thorne74}
\APACinsertmetastar {%
thorne74}%
\begin{APACrefauthors}%
{Thorne}, K\BPBI S.%
\end{APACrefauthors}%
\unskip\
\newblock
\APACrefYearMonthDay{1974}{}{},
\newblock
\unskip
\newblock
\APACjournalVolNumPages{\apj}{191}{}{507-520}.
\PrintBackRefs{\CurrentBib}

\bibitem [\protect \citeauthoryear {%
{Trakhtenbrot}%
}{%
{Trakhtenbrot}%
}{%
{\protect \APACyear {2014}}%
}]{%
trakhtenbrot14}
\APACinsertmetastar {%
trakhtenbrot14}%
\begin{APACrefauthors}%
{Trakhtenbrot}, B.%
\end{APACrefauthors}%
\unskip\
\newblock
\APACrefYearMonthDay{2014}{}{},
\newblock
\unskip
\newblock
\APACjournalVolNumPages{\apjl}{789}{}{L9}.
\PrintBackRefs{\CurrentBib}

\bibitem [\protect \citeauthoryear {%
{Urrutia}%
\ \protect \BOthers {.}}{%
{Urrutia}%
\ \protect \BOthers {.}}{%
{\protect \APACyear {2012}}%
}]{%
urrutia12}
\APACinsertmetastar {%
urrutia12}%
\begin{APACrefauthors}%
{Urrutia}, T.%
, {Lacy}, M.%
, {Spoon}, H.%
, {Glikman}, E.%
, {Petric}, A.%
\BCBL {}\ \BBA {} {Schulz}, B.%
\end{APACrefauthors}%
\unskip\
\newblock
\APACrefYearMonthDay{2012}{}{},
\newblock
\unskip
\newblock
\APACjournalVolNumPages{\apj}{757}{2}{125}.
\PrintBackRefs{\CurrentBib}

\bibitem [\protect \citeauthoryear {%
{Urry}%
\ \BBA {} {Padovani}%
}{%
{Urry}%
\ \BBA {} {Padovani}%
}{%
{\protect \APACyear {1995}}%
}]{%
urry95}
\APACinsertmetastar {%
urry95}%
\begin{APACrefauthors}%
{Urry}, C\BPBI M.%
\BCBT {}\ \BBA {} {Padovani}, P.%
\end{APACrefauthors}%
\unskip\
\newblock
\APACrefYearMonthDay{1995}{}{},
\newblock
\unskip
\newblock
\APACjournalVolNumPages{\pasp}{107}{}{803}.
\PrintBackRefs{\CurrentBib}

\end{thebibliography}

\end{document}